\documentclass[epj]{svjour}
\usepackage{graphicx}
\usepackage{subfig}
\usepackage{revsymb}
\usepackage{amssymb}
\usepackage{color}
\usepackage[normalem]{ulem}

\begin{document}
\title{Dense matter equation of state for neutron star mergers}
\author{S. Lalit \and M. A. A. Mamun \and C. Constantinou \and M. Prakash 
}                     
%
%
\institute{Department of Physics and Astronomy, Ohio University, Athens, Ohio 45701, USA}
%
\date{Received: date / Revised version: date}
%
\abstract{
In simulations of binary neutron star mergers, the dense matter equation of state (EOS) is required over wide ranges of density and temperature as well as under conditions in which neutrinos are trapped, and the effects of magnetic fields and rotation prevail.  
Here we assess the status of dense matter theory and point out the successes and limitations of approaches currently in use. 
A comparative study of the excluded volume (EV) and virial approaches for the $np\alpha$ system using the equation of state of Akmal, Pandharipande and Ravenhall for interacting nucleons is presented in the sub-nuclear density regime. Owing to the  excluded volume of the $\alpha$-particles, their mass fraction vanishes  in the EV approach below the baryon density 0.1 fm$^{-3}$, whereas it continues to rise  due to the predominantly attractive interactions in the virial approach.    
The EV approach of Lattimer et al. is extended here to include clusters of light nuclei such as d, $^3$H and $^3$He in addition to $\alpha$-particles. Results of the relevant state variables from this development  are presented and enable comparisons with related but slightly different approaches in the literature. 
We also comment on some of the sweet and sour aspects of the supra-nuclear EOS. The extent to which the neutron star gravitational and baryon masses vary due to thermal effects, neutrino trapping, magnetic fields and rotation are summarized from earlier studies in which the effects from each of these sources were considered separately. Increases of about $20\% (\gtrsim 50\%)$ occur for rigid (differential) rotation with comparable increases occurring in the presence of magnetic fields only for fields in excess of $10^{18}$ Gauss. Comparatively smaller changes occur due to thermal effects and neutrino trapping. 
Some future studies to gain further insight into the outcome of dynamical simulations are suggested. 
} 


%
\maketitle
%
%
\section{Introduction}
\label{intro}
The first observation of gravitational waves from the merger of binary neutron stars, now known as  GW170817 \cite{GW170817}, has given much impetus to ongoing theoretical investigations of the equation of state (EOS) of dense matter. Analysis of the 
data during inspiral (the phase prior to coalescence) by the LIGO and Virgo collaborations has yielded the chirp mass ${\cal M}=(M_1M_2)^{3/5}/M^{1/5} = 1.188^{+0.004}_{-0.002}~M_\odot$, where $M_1$ and $M_2$ are the companion masses and the total mass $M=M_1+M_2=2.74^{+0.04}_{-0.01}~M_\odot$. Accounting for the component spins in the range inferred from known spinning neutron stars, the individual masses were determined to be in the range $M_1= (1.36-1.60)~M_\odot$ and $M_2= (1.17-1.36)~M_\odot$ when the analysis was restricted to low-spin priors with the dimensionless spin 
$|\chi|=|J_ic/GM_i^2| \ \lesssim 0.05$.  Of particular relevance to the zero-temperature EOS is the limit set by the data on the dimensionless tidal deformability \cite{FH08,F14}
\begin{equation}
\tilde \Lambda = \frac {16}{3} \frac {(M_1+12M_2)M_1^4 \Lambda_1+ (M_2+12M_1)M_2^4 \Lambda_2} {(M_1+M_2)^5} \,.
\end{equation}
For each star, the tidal deformability (or induced quadrupole polarizability) is given by \cite{L1909}
\begin{equation}
\Lambda_{1,2} = \frac 23 k_2 \left( \frac{R_{1,2}c^2}{GM_{1,2}} \right)^5 \,,
\end{equation}
where the dimensionless Love number $k_2$ depends on the structure of star, and therefore on the mass and the EOS. Here, $G$ is the gravitational constant, and $R_{1,2}$ are the radii. The computation of $k_2$ with input EOS's is described in Refs.  \cite{TC67,Hin08,DN09}.  For a wide class of neutron star EOS's, $k_2\simeq$ 0.05-0.15 \cite{Hin08,HLLR10,SPL10}. With unconstrained assumptions about the EOS of each of the stars, Ref. \cite{GW170817} sets the limits $\tilde \Lambda \lesssim 800$ (for low-spin priors with $|\chi| \lesssim 0.05$) and 
 $\tilde \Lambda \lesssim 700$ (for high-spin priors with $|\chi| \lesssim 0.89$). 
 
 Recently, Ref. \cite{De18} has reported results from a reanalysis of the GW170817  data by imposing the common EOS constraint for the structure of both stars. At the $90\%$ credible interval, the bounds $\tilde \Lambda = 222^{+420}_{-138}$ for a uniform component mass prior,  
 $\tilde \Lambda = 245^{+453}_{-151}$  for a component consistent with Galactic double neutron stars, and $\tilde \Lambda = 233^{+448}_{-144}$ for a component mass tallied with the systematics of radio pulsars have been placed.  Across all prior masses, 
a measurement of the common radius in the range $8.9< \tilde R <13.2$ km with a mean value $\langle \tilde R \rangle = 10.8$ km appears to be consistent with the data. This analysis was performed with  polytropic EOS's consistent with constraints from laboratory data up to the nuclear equilibrium or saturation density $n_s\simeq 0.16~{\rm fm}^{-3}$, microscopic calculations of the zero-temperature EOS up to $\sim 2n_s$, and a large number of extrapolations consistent with causality beyond $2n_s$.

 Combining the electromagnetic (EM) \cite{LIGO2017a} and gravitational wave (GW) information from the merger GW170817, Ref. \cite{MM17} provides constraints on the radius $R_{\rm ns}$ and maximum gravitational mass $M^g_{\rm max}$ of a neutron star: 
 \begin{eqnarray}
 M^g_{\rm max} &\lesssim & 2.17 M_\odot \,, \nonumber \\
 R_{1.3} &\gtrsim & 3.1GM^g_{\rm max} \simeq 9.92~{\rm km} \,,
 \end{eqnarray}
 where  $R_{1.3}$ is the radius of a $1.3~M_\odot$ neutron star and its numerical value above corresponds to $M^g_{\rm max}=2.17~M_\odot$.

No evidence of a post-merger signal from GW170817 was found at frequencies up to 4kHZ \cite{GW170817}, the interferometer response at higher frequencies precluding the detection of GW waves exhibiting expected quasi-periodic oscillations of the remnant (see Refs.\cite{BJ12,BSJ16} and references therein).  
Promising prospects for future detections of post-merger signals in upgraded LIGO detectors offer the opportunity to explore the EOS beyond the supra-nuclear densities  afforded by the current data. 
Simulations of the post-merger phase require the EOS of neutron-star matter for wide ranges of physical quantities (See Refs. \cite{BSJ16,fw71,FaRa,MS,BaRe,PaSt}, and references therein).  The baryon number density  ratio $n/n_s$ ranges from $10^{-8}$-$10$,  the latter value depending on the constituents of matter in the core of  a neutron star. Temperatures up to 100 MeV can be reached during the late stages of the merger. Net electron fractions, $Y_e=n_e/n$, ranging from 0.01-0.6, and entropies per baryon S (in units of $k_B$)  in the range 0-100 have been reported in simulations. 

\begin{figure}
\resizebox{0.5\textwidth}{!}{%
    \includegraphics{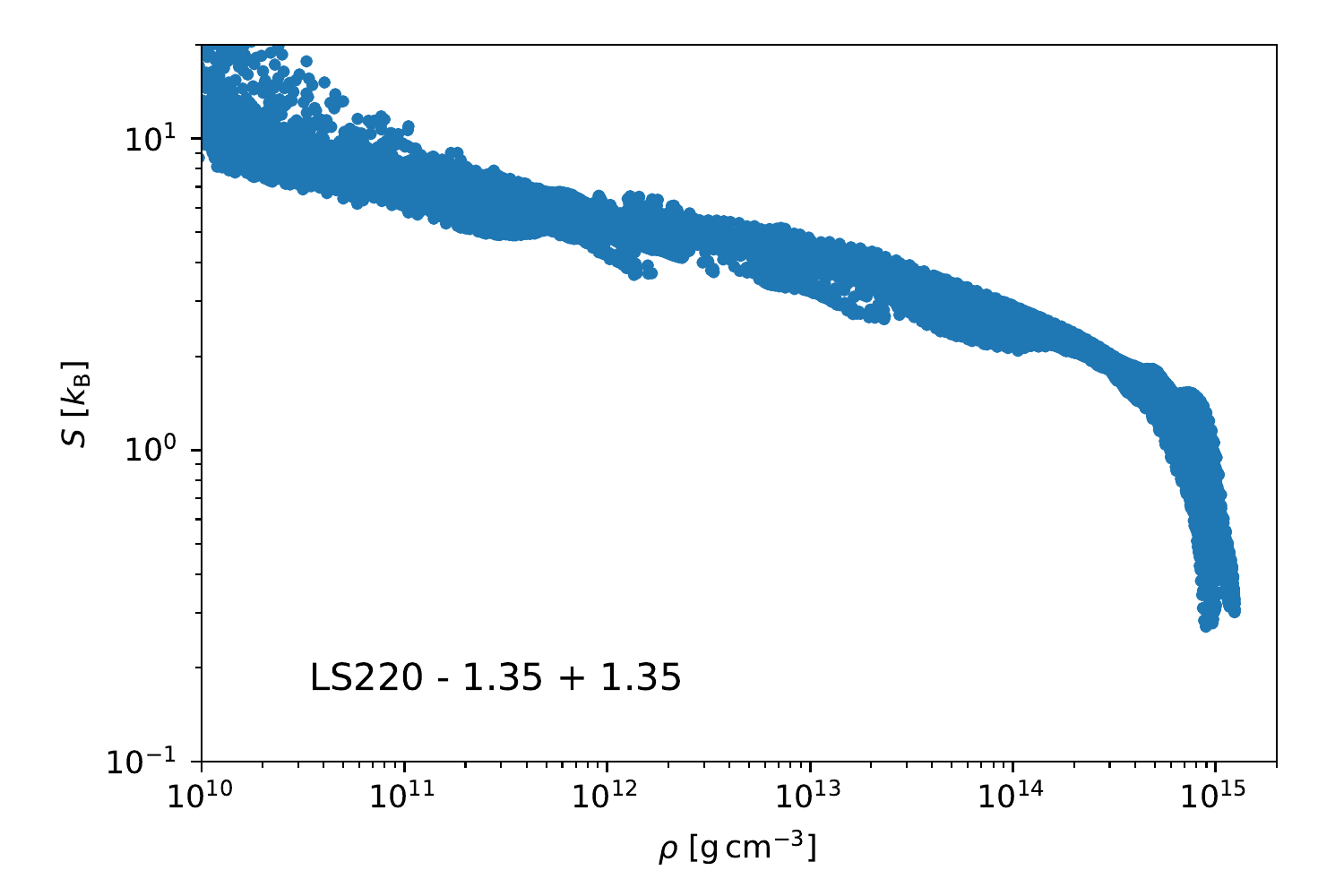}
}
\caption{ Entropy profile of a hyper-massive neuron star during merger. The data represents the meridional plane (i.e., the $x$-$z$ plane, $z$ being the rotational axis). Results are for the nominal EOS of Lattimer and Swesty  (LS220) \cite{LS91} and coalescing neutron stars each of mass 1.35$M_\odot$. Figure courtesy David Radice. }
\label{fig:S}       
\end{figure}

\begin{figure}
\resizebox{0.5\textwidth}{!}{%
    \includegraphics{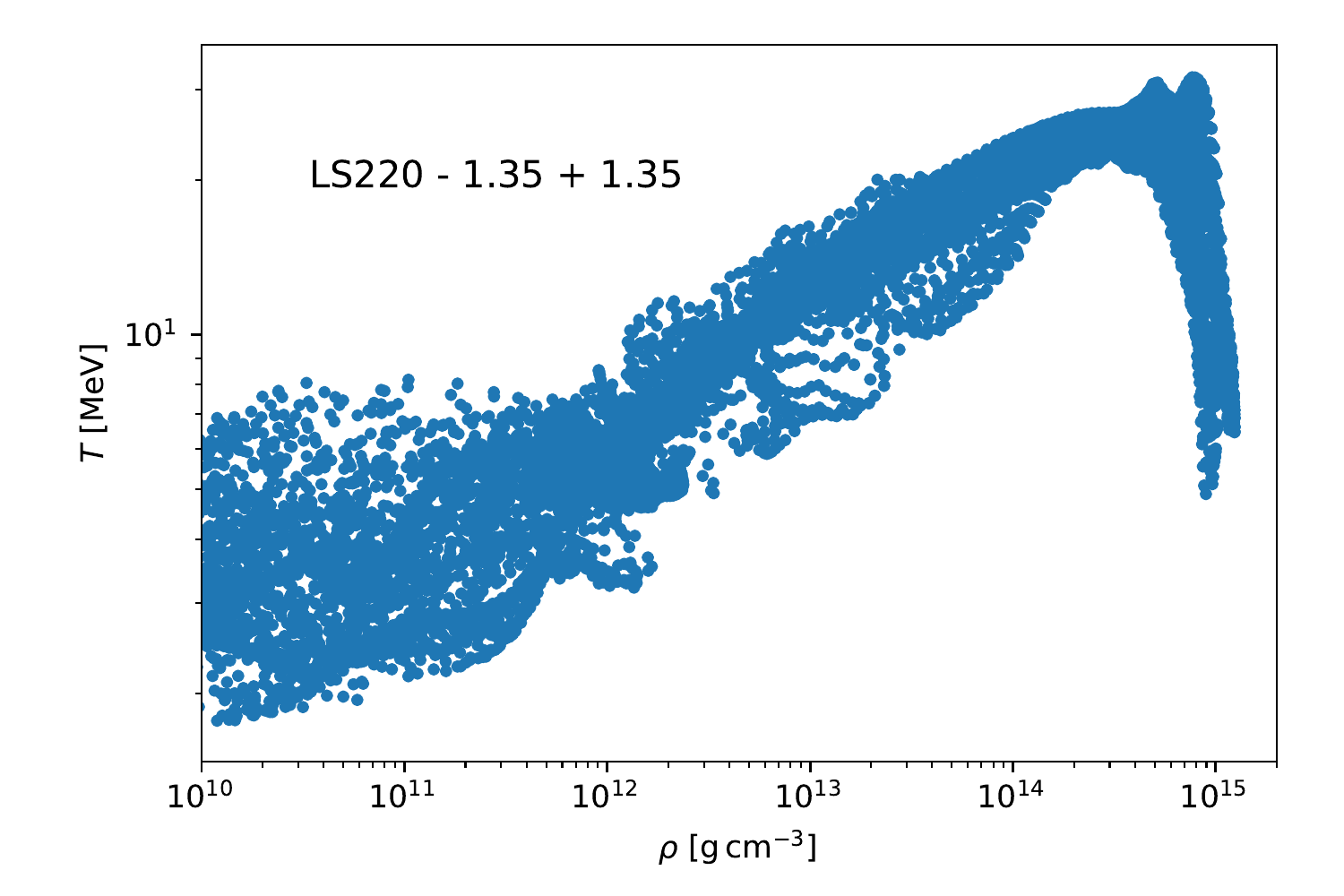}
}
\caption{ Temperature profile of a hyper-massive neuron star during merger.  Details as in Fig. \ref{fig:S}. Figure courtesy David Radice. }
\label{fig:T}       
\end{figure}

Examples of the entropy and temperature profiles vs baryon mass density in the merger of two equal mass neutron stars after about 7.8 ms subsequent to merger are shown in Figs.  \ref{fig:S} and \ref{fig:T}, respectively. Thanks are due to David Radice for providing these results obtained using the EOS of Lattimer and Swesty (LS220) \cite{LS91}. 
The results shown in these figures are drawn from the calculations reported in Refs. \cite{Bern16,Zap18,Rad18}.
The neutron star gravitational masses were 1.35  $M_\odot$ each. The rest-mass of the remnant is approximately 2.7-2.8 $M_\odot$.  Higher values of $S$ and $T$ are attained during the later stages of evolution. A black hole forms $\sim20$ ms after merger in these calculations.

Gravitational signals from the post-merger remnant with a mass close to the maximum gravitational mass, $M^g_{\rm max}$, or even larger for short times, can provide insight into the possible phases of dense matter complementing that offered by electromagnetic signals. Unlike pulsar signals from neutron stars of mass $\gtrsim 2M_\odot$, gravitational signals after coalescence also enable determination of the thermal properties of dense matter. 

One of the main objectives in this paper is  to highlight and  critically assess the sweet and sour spots of the EOS approaches currently in use. The bulk of Sec. \ref{sec:2} is devoted toward this end with new contributions that include a comparison of the excluded volume approach \cite{LS91} using the EOS of APR \cite{APR98} for the np$\alpha$ system with the virial approach \cite{HS06} and an extension of the excluded volume approach to  include  additional light nuclei such as d, $^3{\rm H}$ and $^3{\rm He}$ in addition to $\alpha$-particles. Results of the latter allows for comparisons to be made with earlier works in both of these approaches. Limitations of both of these approaches are also pointed out in this section. 
Another objective is to address the question of  how the masses and radii of neutron stars are affected by thermal effects, composition, trapped neutrinos, magnetic field, and rotation (rigid and differential). Sections \ref{sec:3} through \ref{sec:6} provide brief reviews of earlier works in which  these effects were studied individually along with suggestions for future work that may aid in qualitative and semi-quantitative understanding of the outcome of  dynamical simulations of the mergers of binary neutron stars. Our conclusions are in  Sec. \ref{sec:7}.

%
%

%
\section{Sweet and sour points of the EOS approaches}
\label{sec:2}

Several approaches to the EOS for simulations of core-collapse supernovae, young and old neutron stars, and binary mergers of neutron stars have been developed in  the past decades (see Ref. \cite{OHKT17} for an extensive review). Owing to the different phases of matter encountered at different densities and temperatures, a combination of techniques has been used to calculate the required thermodynamical variables. Broadly speaking, three distinct regions (with different phases and degrees of freedom) in baryon density can be identified: (i) the sub-nuclear density homogeneous and inhomogeneous phases for $n\lesssim 0.1~{\rm fm}^{-3}$, (ii) near-nuclear density homogeneous phase for $0.1<n\lesssim 0.3~{\rm fm}^{-3}$, and (iii) supra-nuclear density with or without phase transitions for $n\gtrsim 0.3~{\rm fm}^{-3}$. 
We use the term ``homogeneous phase'' to refer to a system consisting of hadrons, and leptons of any or all flavors, all regarded as point particles. The same term is also used  for supra-nuclear density matter with or without non-nucleonic degrees of freedom. The term ``inhomogeneous phase'' refers to matter which includes, in addition to nucleons and leptons, composite objects such as light nuclear clusters, heavy nuclei, and pasta-like configurations in which various geometrical shapes [cylindrical (spaghetti), flat (lasagna), cylindrical holes (anti-spaghetti), spherical holes (swiss cheese)] are permitted \cite{RPW83,PR95}. 

As the thermal variables 
depend on $n,Y_e$, and $T$, and on neutrino fractions $Y_{\nu_i}$ when neutrinos of species $i=e,\mu,\tau$ are trapped in matter, the preferred phase will be determined by the minimization of the total free energy with respect to the appropriate variables. As a result, the concentrations of the various species in both phases depend on $(n,Y_e,T)$.  For example, in the range $n/n_s=$ 0.3-0.4 and $Y_e=$ 0.3-0.4 in neutrino-free matter,  nuclei with charge and mass numbers well exceeding 70 and 200 exist  respectively at $T=2$ MeV whereas the corresponding numbers are 30 and 80 at $T=12$ MeV. It must be noted, however, that the precise values depend on the underlying nuclear energy density functionals used in the description of bulk homogeneous matter, nuclei and pasta configurations.  For example, typical values of $T$ below which the pasta phase is present are $\sim 4$ MeV for $Y_e=0.05$ and $\sim 14$ MeV for $Y_e=0.5$ for the EOS of APR \cite{SCMP18}. For charge neutral stellar matter in beta-equilibrium, the dissolution temperature of the pasta phase is around 4-5 MeV (see Ref. \cite{RMRR18} which contains an extensive set of references including classical and quantum molecular dynamical calculations). 

At very low densities, $n\lesssim10^{-6}~{\rm fm}^{-3}$, the abundance of nuclei is generally calculated using the nuclear statistical equilibrium (NSE) approach using mass formulas to calculate the needed chemical potentials $\mu_i$ of nuclei (see \cite{OHKT17} for a review and extensive references). The sour point here is that nuclei not encountered in the laboratory will be present and the use of different mass formulas yields different $\mu_i$ as extrapolations are required. 
In this region, interactions between the nucleons and nuclei are small. 

With the density approaching  $0.1~{\rm fm}^{-3}$ and increasing temperature, however, effects of interactions become progressively important.   Methods devised to account for interactions include the excluded volume approach, the single-nucleus approximation, the full ensemble method, virial expansion etc. \cite{OHKT17}. Matching the NSE results to those of others is also beset with difficulties. Furthermore, the excluded volume approximation lacks attractive interactions, whereas the virial method requires information about phase shifts not always available from experiments in addition to fugacities exceeding unity in certain regions of $n,Y_e,T$ (see below). 

Properties of nucleonic matter in the near-nuclear density region  $0.1<n\lesssim 0.3~{\rm fm}^{-3}$ have received much attention 
recently from  effective field theoretical (EFT) techniques.  Among the advantages of EFT is that systematic error estimates can be made, but the drawback is that it cannot be carried through for densities  $n\gtrsim 0.3~{\rm fm}^{-3}$ due to the perturbative expansion scales reaching invalid regions as the density increases toward the central densities of neutron stars of increasing masses. For the same reason, the exploration of non-nucleonic degrees of freedom such as Bose condensates or quarks is beyond EFT at the current time. 

The discussion above highlights some of the sweet and sour aspects of the current status of dense matter theory.  Clearly, advances in each of the three density regions mentioned above are needed for a fuller microscopic understanding of nuclear matter to better explain  astrophysical phenomena.

\subsection{Instabilities in the sub-nuclear  phase of nucleonic matter}

Our considerations in this subsection are more relevant for the matter produced in intermediate energy heavy ion collisions than for stellar matter discussed in the next subsection. However, the discussion here sets the stage for the case when electrons are present in stellar matter.
A uniform phase of nucleonic matter becomes mechanically unstable (also referred to as spinodally unstable) at sub-nuclear densities, $n< n_s$, for which the compressibility $K(n)=9~dP/dn \leq 0$, where $P=n^2(dE/dn)$ is the pressure. The energy $E$ and pressure of isospin asymmetric matter with $u=n/n_s$ and neutron excess  $\alpha=(n_n-n_p)/n(=n_n+n_p)$ can be written as
\begin{eqnarray}
E(u,\alpha) &=& E(u,0) + \alpha^2S_2(u) + \ldots \\
\frac {P(u,\alpha)}{n_s} &=& u^2 \left[ E^\prime (u,0) + \alpha^2 S_2^\prime(u) \right] + \ldots \,, 
\end{eqnarray}
where the nuclear symmetry energy $S_2(u)= \left. \frac 12 \frac{\partial^2 E(u,\alpha)} {\partial \alpha^2}\right|_{\alpha=0}$ and the prime denotes derivative with respect to $u$. Higher order terms in $\alpha$ are generally small as \\
$S_2(u) >> S_4(u),~S_6(u), \ldots$ (only even powers of $\alpha$ contribute as the two  species are treated symmetrically in the Hamiltonian because of the near-complete isospin invariance of the nucleon-nucleon interaction). As a result, 
\begin{eqnarray}
\frac {\partial (P(u,\alpha)/n_s)}{\partial u} &\simeq& 2u E^\prime(u,0) + u^2 E^{\prime\prime}(u,0)  \nonumber \\ 
&+&  \alpha^2 \left[ 2u S_2^\prime(u,0) + u^2 S_2^{\prime\prime}(u,0) \right] \,. 
\label{dpual}
\end{eqnarray}

An estimate of the density at which the spinodal instability sets in for $T=0$ symmetric nuclear matter (SNM) with $\alpha=0$, or proton fraction $x= (1-\alpha)/2=1/2$,  can be made using the quadratic approximation to the energy vs density close to $u=1$:
\begin{equation}
E(u,0) = -B.E + \frac {K_s}{18} (u-1)^2  \,, 
\label{Evsn}
\end{equation}
where $B.E.=16\pm 1$ MeV is the binding energy of SNM and $K_s=(230\pm 30)$ MeV is its compression modulus. The pressure  $P$ and its density derivative thus become
\begin{equation}
\frac {P}{n_s} = \frac {K_s}{9} u^2 (u-1)\,, \quad \frac {d(P/n_s)}{du} = \frac 13 K_s u \left( u-\frac 23 \right) \,.
\end{equation}
Spinodal instability  sets in at the density
$n_{\rm sp}=(2/3)n_s \simeq 0.11~{\rm fm^{-3}}$ for SNM, independent of $K_S$.  This estimate for $n_{\rm sp}$ is not greatly affected by the skewness of the EOS around $u$, which would add a term 
$\propto (u-1)^3$ to Eq. (\ref{Evsn}).  For the EOS of Akmal, Pandharpande and Ravenhall (APR) \cite{APR98}, $n_{sp} \simeq 0.10~{\rm fm}^{-3}$; other EOS's in current use have similar values of $n_{sp}$.  
With increasing neutron excess ($\alpha \rightarrow 1$), or decreasing proton fraction ($x \rightarrow 0$), the quadratic approximation in $E(u,0)$ or $S_2(u)$ around $u=1$ becomes inadequate \cite{CMPL14}. In this case,  $n_{\rm sp}(u,x)$ can be found from the density ratio $u$ at which Eq.~(\ref{dpual}) vanishes for each $x$.   
For $x \neq 0.5$,  the first and second density derivatives of both $E(u,0)$ and $S_2(u)$ determine $n_{\rm sp}(n,x)$, which decreases from its value for SNM; e.g., $n_{\rm sp}(n,0.1)\simeq 0.05~{\rm fm}^{-3}$ for the EOS of APR.   Other EOS's yield similar qualitative results, quantitative differences being small. 

Thermal effects, which provide positive contributions to the total energy and pressure, also influence the stability of uniform nucleonic matter at sub-nuclear densities.  The onset of the liquid-gas phase transition, determined by the requirements
\begin{equation}
\left. \frac {dP}{dn}\right|_{n_c,T_c} =  \left. \frac {d^2P}{dn^2}\right|_{n_c,T_c} = 0 \,,
\label{spinodal}
\end{equation}
occurs at the critical density $n_c\simeq0.06~{\rm fm}^{-3}$ and critical temperature $T_c\simeq 17.9$ MeV, respectively, for SNM using the EOS of APR \cite{CMPL14}.  Although $T_c/T_c(x=0.5)$ drops significantly with decreasing $x$, the ratio $P_c/(n_cT_c)$ remains very close to 0.347 for $x$ in the range 0.1-0.5 (see Fig. 9 of Ref. \cite{CMPL14}). Qualitative features of the above results  are generic to other EOS's in common use. For example, for the Skyrme EOS parametrization Ska for which $T_c=15.12$ MeV, $P_c/(n_cT_c) = 0.303$, again with little variation in the range of $x=0.1-0.5$ \cite{CMPL14}.

%
\begin{figure}
\resizebox{0.55\textwidth}{!}{%
    \includegraphics{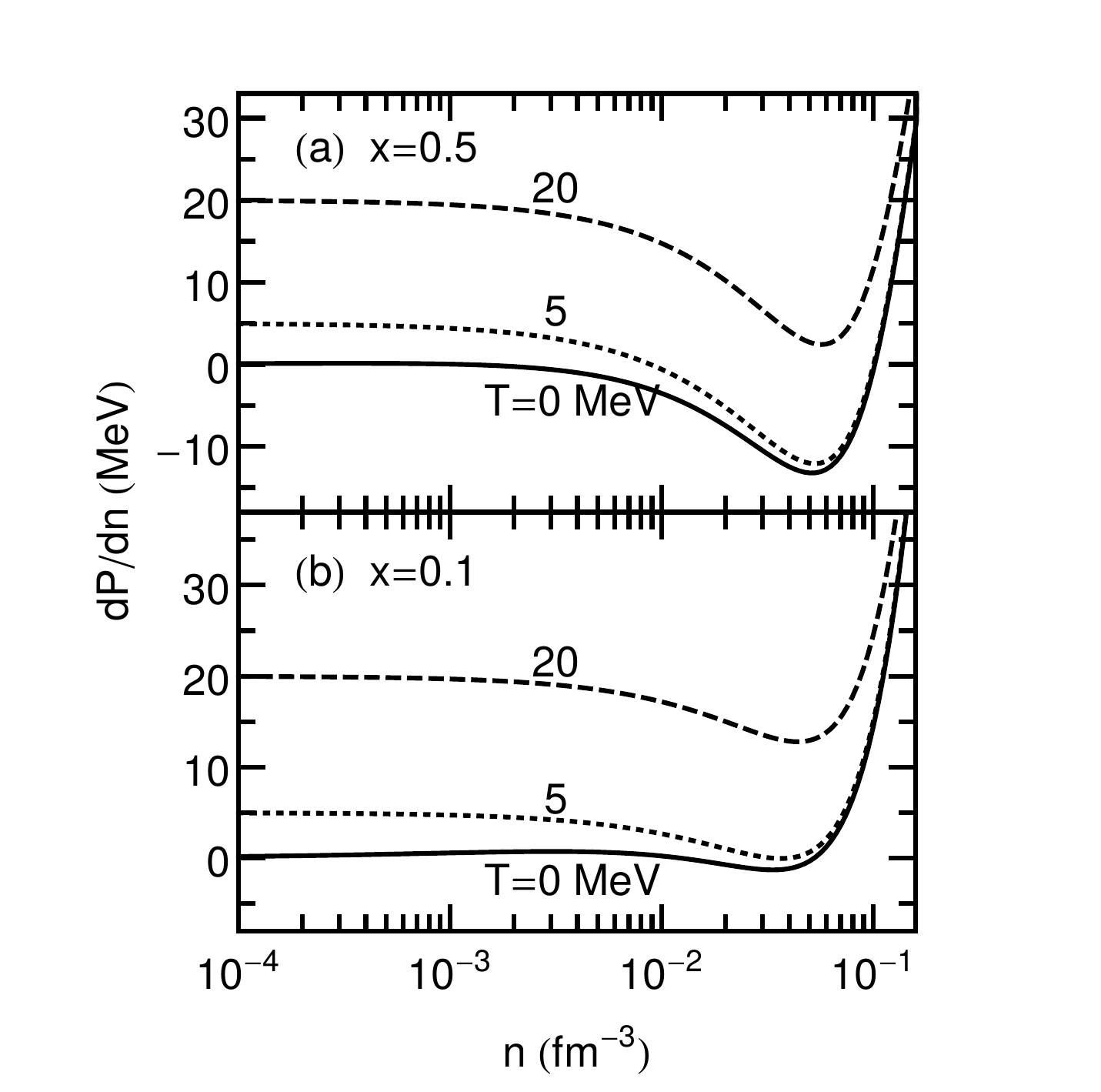}
}
\caption{Derivative of the pressure of nucleons with respect to density vs density at the indicated temperatures $T$ and proton fractions $x$.} 

\label{fig:dPdnT0}       
\end{figure}
%
%

%
\begin{figure}
\resizebox{0.5\textwidth}{!}{%
    \includegraphics{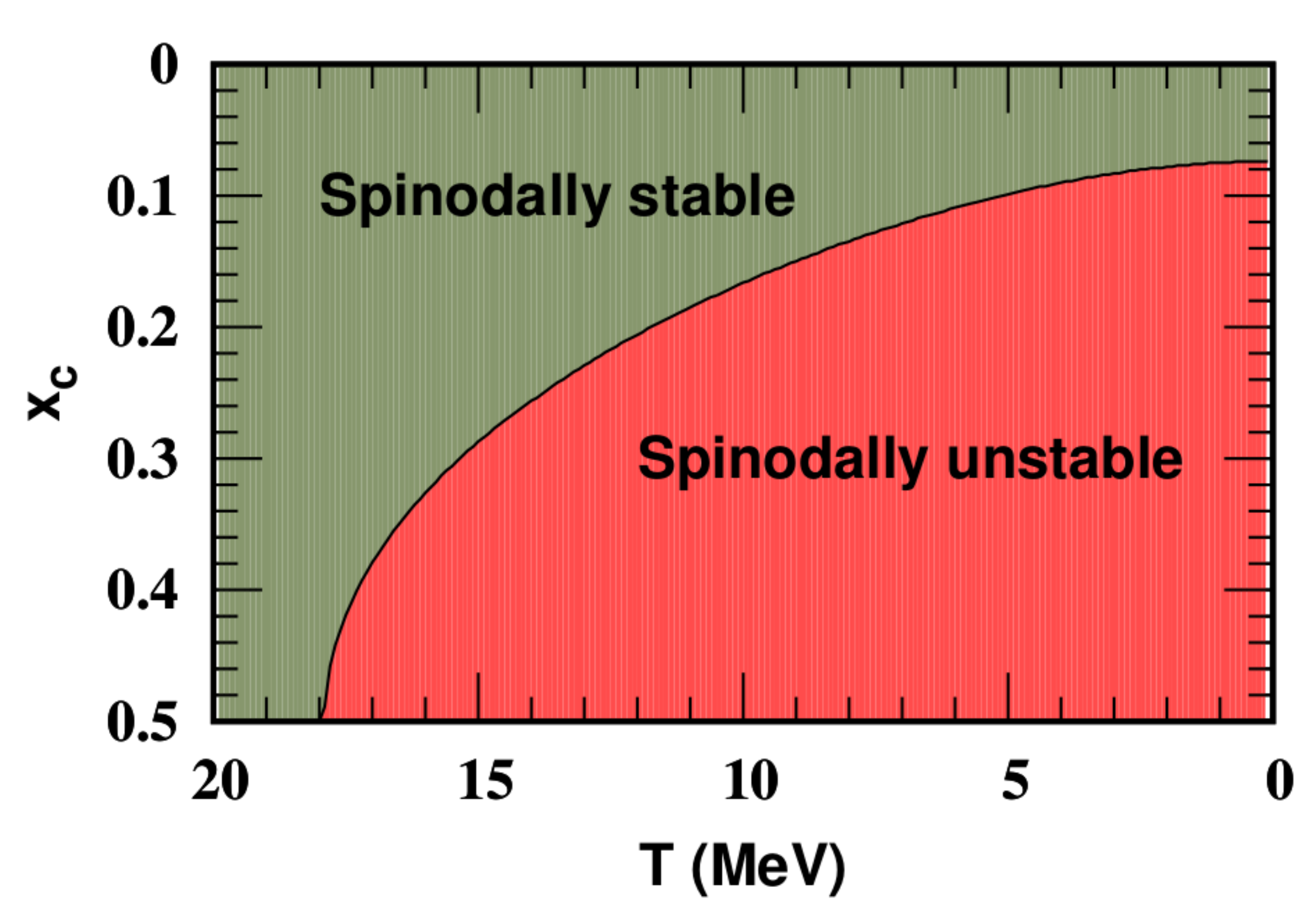}
}
\caption{Spinodally stable and unstable regions in nucleonic matter for the APR model with $x_c$ denoting the proton fraction above which $dP/dn\geq 0$ for all $n$.}
\label{fig:Xc}       
\end{figure}
Figure \ref{fig:dPdnT0} shows the pressure derivative (with respect to density) of the bulk phase of nucleons for different temperatures and proton fractions. In SNM ($x=0.5$), the spinodal region can be clearly identified for the lowest two temperatures shown.  For the same $x$ but at $T=20$ MeV, the spinodal instability is absent as this temperature exceeds the liquid-gas phase transition temperature $T_c=17.91$ MeV for this model. As the proton fraction decreases toward that of pure neutron matter (PNM), the instability region shrinks in size as Fig. \ref{fig:dPdnT0}(b), in which $x=0.1$, shows. Results for intermediate values of $x$ show similar trends \cite{CMPL14}. The value of $x_c$ below which the spinodal density disappears, i.e., for which $dP/dn \geq 0$, is shown as a function of $T$ in Fig. \ref{fig:Xc} with the spinodally stable and unstable regions indicated. For all $T > T_c$, there is no density interval for which spinodal instability occurs as Eq. (\ref{spinodal}) guarantees $dP/dn > 0$ for all $n$. The qualitative behavior of this curve is also exhibited in other models of the EOS  \cite{CMPL14}.

At the densities and temperatures for which the homogeneous uniform phase of nucleonic matter is unstable, an inhomogeneous phase of matter consisting of light nuclear clusters such as   d, $^3{\rm H}, ^3{\rm He}, \alpha$, etc. and heavier nuclei in addition to nucleons becomes energetically favorable. 

\subsection{Stability of the sub-nuclear  phase in stellar matter}

\begin{figure}
\resizebox{0.5\textwidth}{!}{%
\includegraphics{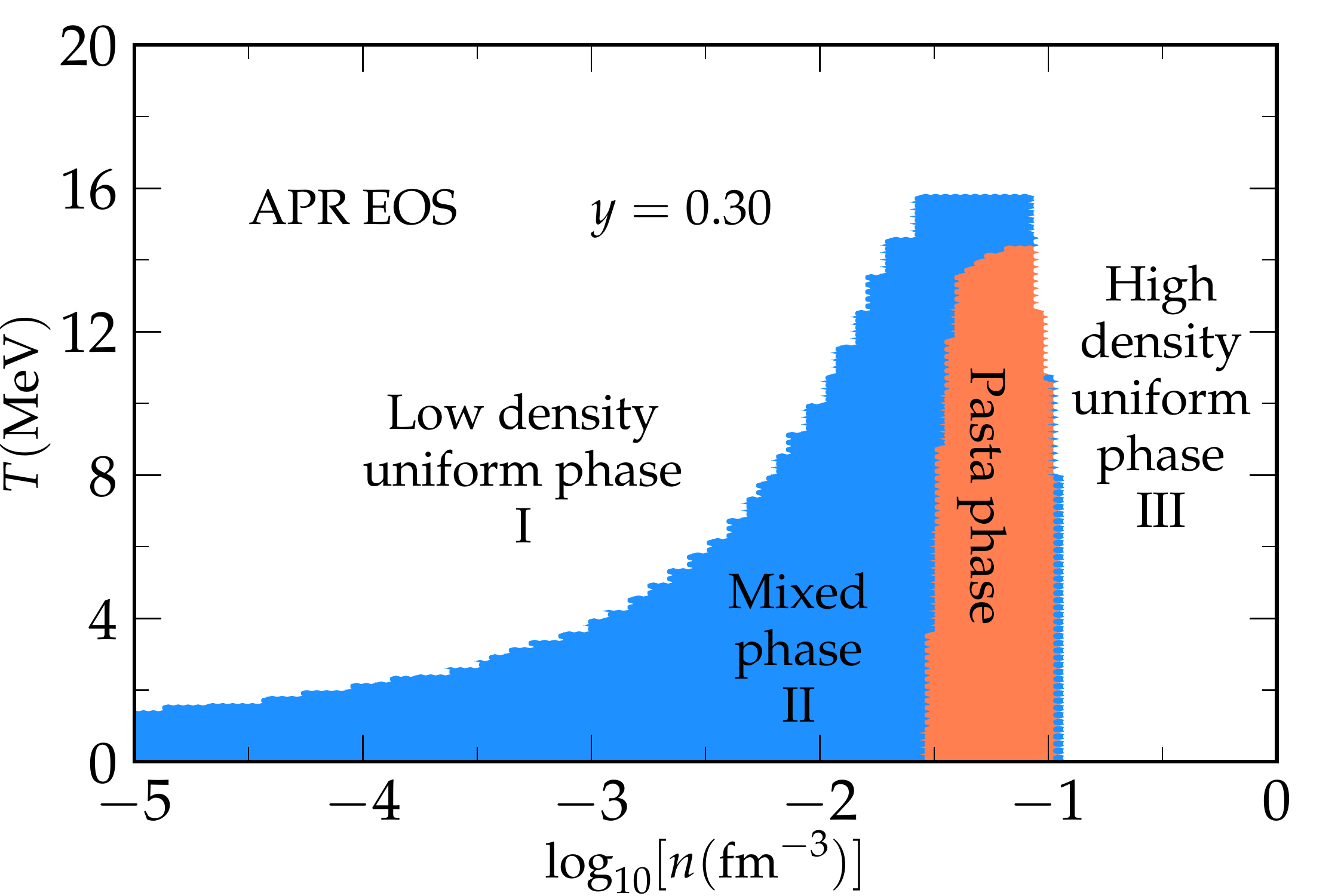}
}
\caption{Phases in dense stellar matter. Figure courtesy   Andre Schneider.}
\label{fig:Schem}       
\end{figure}

Through beta decays and electron capture processes involving  nucleons and nuclei, a uniform background of electrons  
 would also be present in charge neutral stellar matter. The concentration of each species is  determined by the conservation equations for baryon number and charge neutrality together with the minimization of the total free energy with respect to the appropriate internal variables (see below).  Earlier work \cite{LS91,BBP71,LLPR85} has identified three phases of matter depicted in Fig. \ref{fig:Schem}. The generic features  shown in this figure are  for the EOS of APR adopting the excluded volume treatment of Ref. \cite{LS91} for nuclei and for a typical net electron fraction $Y_e$.   Similar features are obtained by using other EOS's and for other $Y_e$'s albeit with quantitative differences.

 In phase I of Fig.  \ref{fig:Schem}, only nucleons, light nuclei \\ (d, $^3{\rm H}, ^3{\rm He}, \alpha$, etc.), and a uniform background of leptons (mostly electrons and positrons, and smaller amounts of muons at high enough $T$'s) to maintain charge neutrality, and photons would be present. Phase II is characterized by the presence of light and heavy nuclei, many rather neutron rich. In a very small region close to $n\simeq 0.1~{\rm fm}^{-3}$, exotic shapes of nuclei commonly called  pasta configurations are also energetically favored. With density increasing beyond  
 $n=0.1~{\rm fm}^{-3}$, the dissolution of all nuclei in the inhomogeneous phase gives way to the uniform phase III of nucleons with charge balancing leptons, and photons.  
 
 At supra-nuclear densities beyond 2-3 $n_s$, matter may also consist of $\Delta$-isobars, Bose (pion, kaon, charged $\rho$-mesons, etc.) condensates, hyperons and/or quarks \cite{MP96}.  At these densities, the effects of baryon superconductivity and superfluidity on the EOS are negligible as the associated gap parameters are small,  $\sim$ 1-2 MeV,  compared to the other energy scales in dense matter. However,  their effects on the long-term cooling of cold-catalyzed neutron stars through neutrino emission are important in interpreting the observed surface temperatures (see the compendium of contributions in Ref. \cite{Page15}). This latter subject is not covered here as it falls outside the scope of this article. \\

%

\subsection*{Electrons restore stability}

\begin{figure}
\resizebox{0.55\textwidth}{!}{%
    \includegraphics{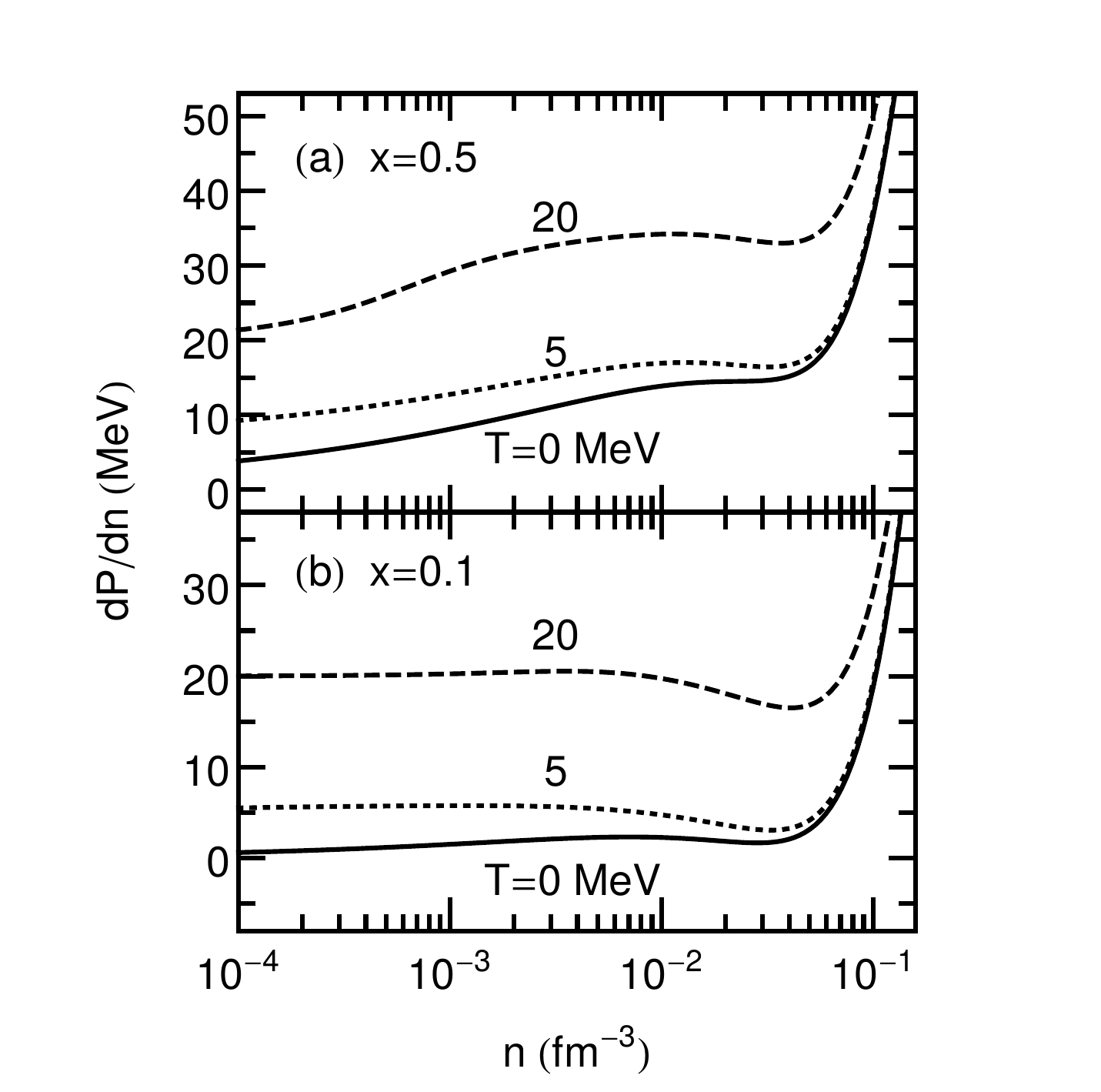}
}
\caption{Derivative of the pressure of nucleons and electrons with respect to density vs density for the indicated temperatures and proton fractions.} 

\label{fig:dPdnT}       
\end{figure}

To highlight the role of electrons in the sub-nuclear phase, we first consider the case in which only nucleons and electrons are present. In this case, the conservation of baryon number and charge neutrality yield $n_n=n(1-x)$ and $n_p=nx=n_e$. 
Minimizing the total free energy density $F(n,x,T)$ with respect to $x$ gives the  energy balance relation between the chemical potentials: 
\begin{eqnarray}
\hat \mu &=& \mu_n-\mu_p = \mu_e \,, \nonumber \\
&\simeq& 4 (1-2x) \left[ S_2(n) + 2 S_4(n) (1-2x)^2 + \cdots \right] \,, 
\end{eqnarray}
which highlights the role of the nuclear symmetry energy. 
The pressure is then found from $P=n^2 \left. \partial (F/n)/\partial n \right|_{x,T}$ from which  the derivative $\left. (\partial P/\partial n)\right|_{x,T}$ can be  calculated. Figure \ref{fig:dPdnT} shows results of this derivative for representative values of $x$ and $T$ as functions of $n$. The results here were calculated without approximation using the Hamiltonian density of APR for the nucleons with electrons being treated as a free Fermi gas.  

%

It is clear from Fig. \ref{fig:dPdnT} that contributions from electrons to the total pressure {\it entirely remove the mechanical (spinodal) instability} present in baryons-only matter for all $x$ and $T$. A similar conclusion was reached by Ref. \cite{CMPL15} in which the adiabatic index $\Gamma_S = \left. (n/P)(\partial P/ \partial n)\right|_S$, where $S$ is the entropy per baryon,  was calculated (in Sec. VI there) for other nonrelativistic models (MDYI and SkO$^\prime$) and a mean field  theoretical model of nucleons. This conclusion also applies for charge neutral and beta-stable neutron star matter at both zero and finite temperatures for which the equilibrium proton fraction $\tilde x$ varies with $n$. In this case, the requirement that pressure be a 
continuously increasing function  of $n$ in a stable star assures stability against spinodal collapse. Note also that baryons-only matter is mechanically stable at all $T$ for roughly $x<0.1$. For these proton fractions, electrons are required only for the purposes of charge neutrality.

\subsection{Inclusion of light nuclear clusters}

Although electrons restore mechanical stability in stellar matter, the presence of light nuclear clusters such as $d, ^3{\rm H}, ^3{\rm He}, \alpha$, etc. lowers the free energy and thus becomes the favored state of matter in phase I of Fig. \ref{fig:Schem}.  In what follows, we discuss the excluded volume and virial approaches commonly used for treating the presence of these light nuclear clusters. An extension of the excluded volume approach of Refs. \cite{LS91,LLPR85} to include other clusters beyond $\alpha$-particles will be presented. A comparison of the excluded volume and virial approaches for the np$\alpha$ system with leptons and photons will be made together with a brief discussion of the advantages and drawbacks of each of these approaches.

\subsection*{The excluded volume approach with $\alpha$- particles}

With a binding energy $B_\alpha\simeq 28.3$ MeV, $\alpha$-particles are the most bound among the light nuclear clusters.
The discussion here to include $\alpha$- particles in addition to nucleons, electrons, and photons follows closely that of Ref. \cite{LS91} which, in turn, is  a simplified version of an earlier treatment in Ref. \cite{LLPR85}. We first outline the procedure of Ref. \cite{LS91} here as it paves the way for the subsequent inclusion of additional light nuclear clusters in this approach.

 Interactions between $\alpha$-particles and nucleons (assumed to be point particles) are taken into account by treating the  $\alpha$-particles as rigid spheres of  volume \\ $v_\alpha = \frac {4\pi}{3}(\frac 45 a_{{\rm p}\alpha})^3\simeq 24~{\rm fm}^3$, where  $ a_{{\rm p}\alpha}$ is the proton-$\alpha$ scattering length. This treatment accounts for repulsive interactions only, attractive interactions being deemed as small. The conservation equations for baryon number and charge neutrality are
 \begin{eqnarray}
 \label{bcons}
 n &=& 4n_\alpha + ( n_{n{\rm o}} +  n_{p{\rm o}} ) (1-n_\alpha v_\alpha) \\
 nY_e &=& 2n_\alpha + n_{p{\rm o}} (1-n_\alpha v_\alpha) \,,
 \label{ccons}  
 \end{eqnarray}
 where $n_{n{\rm o}}$ and $n_{p{\rm o}}$ are the neutron and proton densities outside the $\alpha$-particles of density $n_\alpha$,  and $Y_e=n_e/n$ is the net electron fraction (i.e., $n_e=n_{e^-} - n_{e^+}$). Equation (\ref{bcons}) enables the mass fractions to be defined as
\begin{eqnarray}
X_\alpha &=& 4 \frac {n_\alpha}{n}\,, \nonumber \\
X_{n{\rm o}} &=& \frac {n_{n{\rm o}} (1-n_\alpha v_\alpha)}{n} \,,~~
X_{p{\rm o}} = \frac {n_{p{\rm o}} (1-n_\alpha v_\alpha)}{n} \,. 
\end{eqnarray}
The total free energy density can be decomposed as 
 \begin{equation}
 F = F_b + F_e + F_\gamma \,,
 \end{equation}
 where $F_b=F_{\rm o} + F_\alpha$ is the free energy density of  baryons, $F_e$ and $F_\gamma$ are those of the leptons and photons. The component $F_{\rm o}$ refers to the outside (of $\alpha$-particles) nucleons and can be written as 
 \begin{equation}
 F_{\rm o} = (1-n_\alpha v_\alpha)~ n_{\rm o} f_{\rm o} (n_{\rm o}, x_{\rm o},T) \,, 
 \end{equation}
where $f_{\rm o}$ is the free energy per nucleon, $n_{\rm o} =  n_{n{\rm o}} +  n_{p{\rm o}}$ and $x_{\rm o} =  n_{p{\rm o}}/n_{\rm o}$.  
The quantity $f_{\rm o}$ can be calculated using a suitable model for the EOS of interacting nucleons. Here we use that of APR at finite $T$ following Ref. \cite{CMPL14} where details for calculating $F_e$ and $F_\gamma$ are also provided. The $\alpha$-particles are treated as non-interacting Boltzmann particles, whence  
\begin{equation}
F_\alpha = n_\alpha (\mu_\alpha  - B_\alpha -T) \,.
\label{Fal} 
\end{equation}
The $\alpha$-particle chemical potential, easily obtained from the classical gas expression $n_\alpha = 8n_Q \exp(\mu_\alpha/T)$,  is
\begin{equation}
\mu_\alpha = T \ln \left( \frac {n_\alpha}{8n_Q} \right)  \quad {\rm with} \quad  n_Q = \left( \frac {mT}{2\pi \hbar^2} \right)^{3/2} \,,
\label{mual}
\end{equation}
where $n_Q$ is the quantum concentration of nucleons with $m$ denoting the nucleon mass. Being non-interacting particles, the $\alpha$-particle pressure, energy density    and entropy density are 
\begin{eqnarray}
P_\alpha &=& n_\alpha T\,, ~~\epsilon_\alpha = n_\alpha \left( \frac 32  T -B_\alpha \right)\,, ~~{\rm and} \nonumber \\
s_\alpha &=& n_\alpha \left( \frac 52 - \frac {\mu_\alpha}{T} \right) \,.
\label{Pesal}
\end{eqnarray}
Minimization of $F$ with respect to $n_\alpha$ yields the relationship between the chemical potentials of the baryons:
\begin{equation}
0 = \frac {\partial F}{\partial n_\alpha} \Rightarrow  \mu_\alpha = 2( \mu_{n{\rm o}} + \mu_{p{\rm o}} ) + B_\alpha - p_{\rm o} v_\alpha \,,
\label{mus}
\end{equation}
where $p_{\rm o}$ is the pressure of nucleons outside the $\alpha$-particles obtained using the (purely nucleonic) EOS at the given (subnuclear) density, lepton fraction, and temperature. Because the EOS of APR uses a common value for the rest masses for the neutron and proton, the term involving $2(m_p-m_n)$ is not included in the above equation.   

In simulations of core-collapse supernovae, proto-neutron stars and binary mergers of neutron stars, the EOS is generally tabulated in terms of the variables $(n, Y_e,T)$. From a numerical standpoint, it is advantageous to extend the variables to $(n,Y_e,X_p,T)$, where $X_p = n_{p{\rm o}}/n$. Eliminating $n_\alpha$ from Eqs. (\ref{bcons}) and (\ref{ccons}), one obtains
\begin{eqnarray}
  n_o &=& \frac{2n(1-2Y_e)+nX_p(4-nv_{\alpha})}{2-nY_ev_{\alpha}} \, \nonumber \\ 
    n_{po} &=& nX_p\,, \quad
  n_{no} = n_o - n_{po} \,, ~~{\rm and}~~
  x_o =\frac{n_{po}}{n_o} \,.
\end{eqnarray}
Thus $n_{po}$ and $n_{no}$ are completely determined by specifying $(n, Y_e, T)$ and providing a guess value for $X_p$. With these $n_{po}$ and $n_{no}$ at hand, $\mu_{po}$ and $\mu_{no}$ [needed for the calculation of $n_{\alpha}$ using Eq. (\ref{mual})] can be obtained by solving  
\begin{equation}
2\int \frac{d^3p}{(2\pi\hbar)^3}\frac{1}{1+\exp\left(\frac{e_i -\mu_i}{T}\right)} - n_i =0
\end{equation}
where the spectra $e_i$ correspond to the model of choice for the nucleonic EOS. 
The value of $X_p$ can then be updated iteratively to satisfy the baryon number conservation Eq. (\ref{bcons}).
The total pressure, entropy density and energy density are then 
\begin{eqnarray}
P &=& P_{\rm o} + P_\alpha + P_e + P_\gamma \,, \nonumber \\
s &=& (1-n_\alpha v_\alpha) s_{\rm o} + s_\alpha + s_e + s_\gamma \, \quad S=s/n \nonumber \\
\epsilon &=& \sum_i \mu_i n_i + Ts - P\,, \quad E = \epsilon /n \,,
\end{eqnarray}
where $S$ denotes the total entropy per baryon, $E$ the total energy per baryon and $P_o = (1-n_\alpha v_\alpha)~p_o$. In utilizing the thermodynamic relation above to obtain the energy density, it is necessary to account for the $\alpha$-particle  binding energy $B_\alpha$ in the total chemical potential of the $\alpha$, i.e., $\mu_\alpha^{\rm tot} = \mu_\alpha - B_\alpha$. We defer presentation of the numerical results of  these state variables to a later section.

\subsection*{The excluded volume approach with multiple clusters}
The presence of additional light nuclear clusters such as d, $^3$H, and $^3$He give a lower free energy relative to the case when only $\alpha$-particles are considered besides nucleons. The binding energies of these light nuclei are listed in Table \ref{tab:0}.  In the excluded volume approach, interactions between the various nuclear species and nucleons can also be included by treating these nuclei as rigid spheres with  excluded volumes $v_i=\frac 43 \pi R_i^3$, where the sharp sphere radii $R_i$ can be inferred from the measured charge or mass rms radii, the latter not being experimentally available as yet. Values of $v_i,~i =$ d, $^3$H, $^3$He and $^4$He are also listed in Table \ref{tab:0}. Note that $v_\alpha$ differs slightly from that used in Refs. \cite{LS91,LLPR85}. 

\begin{table}[hbt]
\caption{Properties of light nuclei. The symbol for each nucleus used in text is as indicated. The symbol $v$ stands for the rigid sphere effective excluded volume.}
\label{tab:0}       
\begin{center}
\begin{tabular}{cccc}
\hline\noalign{\smallskip}
Nucleus  & Symbol &   B.E. (MeV) & $v~({\rm fm}^3)$    \\ \hline
d & d & 2.22 & 40.5 \\
$^3$H & $\tau$ &  8.48 & 23.2 \\
$^3$He &  3 & 7.72 & 32.0 \\
$^4$He & $\alpha$ & 28.3 & 19.9 \\
\hline\noalign{\smallskip}
\end{tabular}
\end{center}
\end{table}

The baryon number conservation equation takes the form 
\begin{eqnarray}
n &=& 4n_{\alpha} {+} (1-n_{\alpha}v_{\alpha}) \nonumber \\
&\times& \left\{ 3n_3+(1-n_3v_3) \right. \nonumber \\
&\times& \left[ 3n_{\tau}+(1-n_{\tau}v_{\tau}) \right. \nonumber \\
&\times& \left. \left. \left( 2n_d+n_o(1-n_dv_d)\right)\right]\right\} \,.
\label{Bcons}
\end{eqnarray}
Note that the order by which the excluded volumes are nested does not affect the final results for the thermodynamic state variables as functions of $(n,Y_e,T)$ insofar as a particular order is used consistently over the course of the calculation.  

The mass fractions are defined as 
\begin{eqnarray}
X_\alpha &=&  \frac {4n_\alpha}{n}\,, X_3=\frac {3n_3}{n} (1-n_\alpha v_\alpha) \,,  \nonumber \\
X_\tau &=&  \frac {3n_\tau}{n} (1-n_\alpha v_\alpha) (1-n_3 v_3) \,,  \nonumber  \\
X_d &=&  \frac {2n_d}{n} (1-n_\alpha v_\alpha) (1-n_3 v_3) (1-n_\tau v_\tau) \,,  \nonumber  \\
X_{n{\rm o}} &=&  \frac {n_{n{\rm o}}} {n} \prod_i (1-n_i v_i) \,,
X_{p{\rm o}} = \frac {n_{p{\rm o}}} {n} \prod_i (1-n_i v_i) \,. 
\label{MMfracs}
\end{eqnarray}
with $i=d,3,\tau,\alpha$.
The corresponding charge neutrality condition requires
\begin{eqnarray}
 nY_e &=& 2n_{\alpha} + (1-n_{\alpha}v_{\alpha}) \nonumber \\
 &\times& \left\{2n_3+(1-n_3v_3) \right. \nonumber \\
 &\times&
\  \left[n_{\tau}+(1-n_{\tau}v_{\tau}) \right. \nonumber \\
&\times& \left. \left. \left(n_d+n_{po}(1-n_dv_d)\right)\right]\right\} \,.
\label{Ccons}
\end{eqnarray}
When all of the nuclear species are treated as non-interacting Boltzmann particles, their densities, chemical potentials and ``bare'' (i.e., without excluded volume factors included) free energy densities are given by
\begin{eqnarray}
\label{Dens}
n_i &=& A_i^{3/2}n_Q \exp(\mu_i/T) \nonumber \\ 
\mu_i &=& T \ln \left( \frac{n_i}{A_i^{3/2} n_Q} \right)  \nonumber \\
\label{Chems}
f_i &=& n_i(\mu_i  - B_i -T) \quad i=d,\tau,3,\alpha \,,
\label{Fis}
\end{eqnarray} 

\noindent where $A_i$ are the mass numbers of the light nuclei and $n_Q$ is the quantum concentration of nucleons in Eq. (\ref{mual}). 
The pressure, energy density and entropy density of each species is that of a non-interacting gas as in Eq. (\ref{Pesal}).
As in the previous section, the free energy densities of interacting nucleons outside the nuclei are calculated using the EOS of APR at finite $T$ \cite{CMPL14}. 

The relationships between the various chemical potentials are obtained by 
minimizing the total free energy density
\begin{eqnarray}
F &=& F_b + F_e + F_\gamma \nonumber \\
F_b &=& F_{{\rm o}} + F_d + F_\tau + F_3 + F_\alpha \nonumber \\
F_{\rm o} &=& \prod_{i=d,\tau,3,\alpha} (1-n_iv_i)~ n_{\rm o} f_{\rm o} (n_{\rm o}, x_{\rm o}, T) \nonumber \\
F_{\rm d} &=& \prod_{i=\tau,3,\alpha} (1-n_iv_i)~  f_{\rm d} (n_{\rm d}, T) \nonumber \\
F_{\rm \tau} &=& \prod_{i=3,\alpha} (1-n_iv_i)~  f_{\rm \tau} (n_{\rm \tau}, T) \nonumber \\
F_{\rm 3} &=& (1-n_{\alpha}v_{\alpha})~ f_{\rm 3} (n_{\rm 3}, T) \nonumber \\
F_{\rm \alpha} &=& f_{\rm \alpha} (n_{\rm \alpha}, T)
\end{eqnarray}
\noindent with respect to $n_i$:
\begin{eqnarray}
\frac {\partial F}{\partial n_i} &=& 0 \Rightarrow \nonumber \\
  \mu_d &=& \mu_{no}+\mu_{po}+B_d-v_dp_o \nonumber \\
  \mu_{\tau} &=& 2\mu_{no}+\mu_{po}+B_{\tau}-v_{\tau}(p_o+n_dT) \nonumber \\
  \mu_3 &=& \mu_{no}+2\mu_{po}+B_3-v_3(p_o+n_dT+n_{\tau}T) \nonumber \\
  \mu_{\alpha} &=& 2(\mu_{no}+\mu_{po})+B_{\alpha}-v_{\alpha}(p_o+n_dT+n_{\tau}T+n_3T) \,, \nonumber \\
\label{chems}
\end{eqnarray}
where $p_{\rm o}$ stands for the pressure of interacting nucleons in the absence of the light nuclei.
The inclusion of additional clusters increases the number of quantities to be determined compared to the case when only $\alpha$-particles are considered.  As in the previous section, the set of variables $(n,Y_e,X_p,T)$ facilitates numerical evaluations considerably.  The nucleon densities outside the nuclei can be found by eliminating $n_\alpha$ from Eqs. ({\ref{Bcons}) and ({\ref{Ccons}) and a guess value of $X_p$: 
\begin{eqnarray}
  n_o &=& \left\{\left[ Q \frac{1}{1-n_3v_3}-3n_{\tau}\right]\frac{1}{1-n_{\tau}v_{\tau}}-2n_d\right\}
    \frac{1}{1-n_dv_d}  \nonumber \\
    Q &=& \left(\frac{2n(1-2Y_e)+R(4-nv_{\alpha})}{2-nY_ev_{\alpha}}-3n_3\right) \nonumber \\
      R &=& 2n_3+(1-n_3v_3) \nonumber \\
      &\times& \left\{n_{\tau}+(1-n_{\tau}v_{\tau})\left[n_d+nX_p(1-n_dv_d)\right]\right\} \nonumber \\
        n_{no} &=& n_o - n_{po} \,,  \quad 
  x_o = \frac{n_{po}}{n_o} \quad   {\rm and} \quad n_{po} = nX_p \,
\end{eqnarray}
using which the outside nucleon chemical potentials can be determined by utilizing their relations to the nucleon densities. 
Equation (\ref{chems})  then provides the various chemical potentials and Eq. (\ref{Chems}) their corresponding densities. The correct value of $X_p$ can be determined iteratively by satisfying the baryon number conservation Eq. (\ref{Bcons}).
The total pressure is then given by 
\begin{eqnarray}
P &=& P_b + P_e + P_\gamma \nonumber \\
P_b &=& P_{{\rm o}} + P_d + P_\tau + P_3 + P_\alpha \nonumber \\
P_{\rm o} &=& \prod_{i=d,\tau,3,\alpha} (1-n_iv_i)~ p_{\rm o} (n_{\rm o}, x_{\rm o}, T) \nonumber \\
P_{\rm d} &=& \prod_{i=\tau,3,\alpha} (1-n_iv_i)~  p_{\rm d} (n_{\rm d}, T) \nonumber \\
P_{\rm \tau} &=& \prod_{i=3,\alpha} (1-n_iv_i)~  p_{\rm \tau} (n_{\rm \tau}, T) \nonumber \\
P_{\rm 3} &=& (1-n_{\alpha}v_{\alpha})~ p_{\rm 3} (n_{\rm 3}, T) \nonumber \\
P_{\rm \alpha} &=& p_{\rm \alpha} (n_{\rm \alpha}, T) \nonumber \\
p_i &=& n_i T
\label{MPres}
\end{eqnarray}
Expressions for the entropy and energy densities are similar with $s_{\rm o}$ and $\epsilon_{\rm o}$ taking the place of 
$p_{\rm o}$, and  the various other $p_i$'s above replaced by $s_i = n_i (5/2 - \mu_i/T)$ and $\epsilon_i = n_i (3/2T-B_i)$ for  $i=d,\tau,3,\alpha$, respectively. 

\subsection*{The virial approach}

The treatment of clusters is also afforded by the virial expansion approach that includes bound and continuum states, and provides corrections to the ideal gas results for thermal variables \cite{BU37}.   When applicable, this approach is model independent as experimental data (i. e., phase shifts), where available, are input to theory. In terms of the partition function ${\cal Q}$, the pressure $P = (T/V)\log {\cal Q}$, and  is expressed in terms of  
the fugacities $z_i = \exp (\mu_i/T)$,  ($i =$ d, $^3$H, $^3$He, $^4$He etc.) and the 2nd virial coefficients $b_2$ which are simple integrals involving thermal weights and elastic scattering phase shifts.  Unlike in classical statistical mechanics, a satisfactory treatment of the 3rd and higher order quantum virial coefficients is yet to be accomplished.  

In the context of heavy-ion collisions, this approach was used to calculate the state variables of a resonance gas in Ref. \cite{VP92}. 
For low-density supernova and neutron-star matter, the EOS for the np$\alpha$ system was calculated in Ref. \cite{HS06}, with the inclusion of other light nuclear clusters in Refs. \cite{OGHSB07,Arcones08} and those with heavy-nuclei also in Ref. \cite{SHT10}. The review in  Ref. \cite{OHKT17} provides an extensive list of other references in which the virial approach and its extensions have been used to calculate the EOS of low-density matter. 

Here we restrict ourselves to the np$\alpha$ system with electrons and photons to make comparisons with the results of the excluded volume approach, and to point out similarities and differences.
The results reported here were obtained precisely in the manner expounded in Refs. \cite{HS06} and \cite{LS91} respectively, with APR serving as the underlying model for the nucleon-nucleon force in the latter case.
   In both cases, we have verified that our results shown here reproduce those of these earlier works.

\subsection*{Comparison between the excluded volume and virial approaches}

We begin by showing results for the np$\alpha$ system  with leptons and photons. 
Figure \ref{fig:Xas} shows a comparison of the $\alpha$-particle mass fractions $X_\alpha=4n_\alpha/n$ between the excluded volume approach using the EOS of APR for nucleons and the virial approach. Unlike in the standard LS model, $X_\alpha$'s here lie above the virial results until they vanish at some density (see Fig. 5 and associated discussion in Ref. \cite{HS06}). 
Clearly, results of the excluded volume approach depend on the treatment of  the nucleon-nucleon interactions. As noted in Refs. \cite{LS91,LLPR85}, the $p_{\rm o}v_\alpha$ term in Eq. (\ref{mus}) decreases the $\mu_\alpha$ and $n_\alpha$ as the density increases causing the disappearance of $\alpha$'s. Such is not the case in the virial approach in which for each $Y_e$, $X_\alpha$'s continue to increase up to and even beyond $n_s$ where the approach becomes invalid. However, the appearance of heavy-nuclei at sub-nuclear densities results in diminishing $X_\alpha$'s in the virial approach as well \cite{SHT10}. The physical difference between the two approaches is that only repulsive interactions are incorporated in the excluded volume approach whereas in the virial approach the  $\alpha$-nucleon interactions are predominantly attractive.  In what follows, we will first present results from each of these approaches to provide a comparison and to appreciate their characteristics before addressing  a method in which the limitations of each method can be avoided.

\begin{figure}
\centering
\subfloat{\includegraphics[width=0.55\textwidth,keepaspectratio]{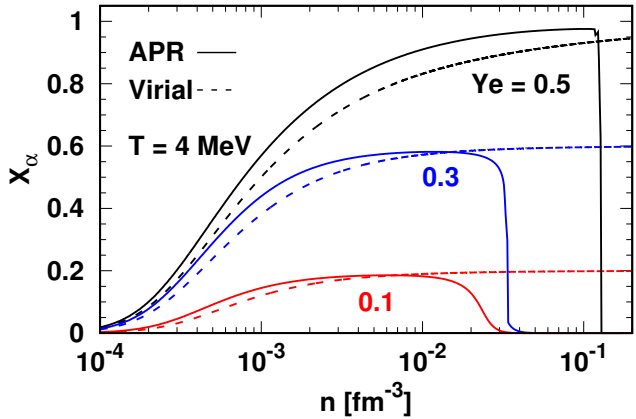}}
\vspace{-1.0cm}
\caption{Mass fractions $X_\alpha=4n_\alpha/n$ vs baryon density at the indicated temperature $T$ and net electron fractions $Y_e$.} 
\label{fig:Xas}
\end{figure}

Figures \ref{fig:Ps}(a) and (b)  show the pressure vs baryon density in the two approaches at $T=4$ MeV, and for $Y_e=0.1$ and 0.4, respectively. The individual contributions from the baryons presented in this figure provide a contrast between results of  the excluded volume  approach (solid curves) using  the EOS of  APR for nucleons and the virial (dashed curves) approach.   Unlike in the virial approach, the excluded volume pressure due to the outside nucleons, $P_{\rm o}$,  shows a non-monotonic behavior for both $Y_e$'s. This difference is due to the disappearance of the $\alpha$-particles with growing density in the excluded volume approach.  Particularly noteworthy are the negative values of $P_{\rm o}$ for $Y_e=0.4$ characteristic of nearly symmetric nucleonic matter at the densities shown \cite{CMPL14}. This feature is absent in the virial approach. The sub-dominant contributions $P_\alpha$ from the $\alpha$-particles are nearly the same in the two approaches.  For both $Y_e$'s, the contribution from the leptons, $P_e$,  is dominant over that of the baryons. 
This dominance persists for all values of $Y_e$'s except  those approaching that of PNM for which $Y_e=0$. The contribution from photons is negligible at the temperature and densities shown.  We note, however, that these results, along with those of other state variables to be shown below,  will be quantitatively altered when other light nuclear clusters as well as heavy nuclei are included at sub-nuclear densities.

\begin{figure}
\centering
\subfloat{\includegraphics[width=0.5\textwidth,keepaspectratio]{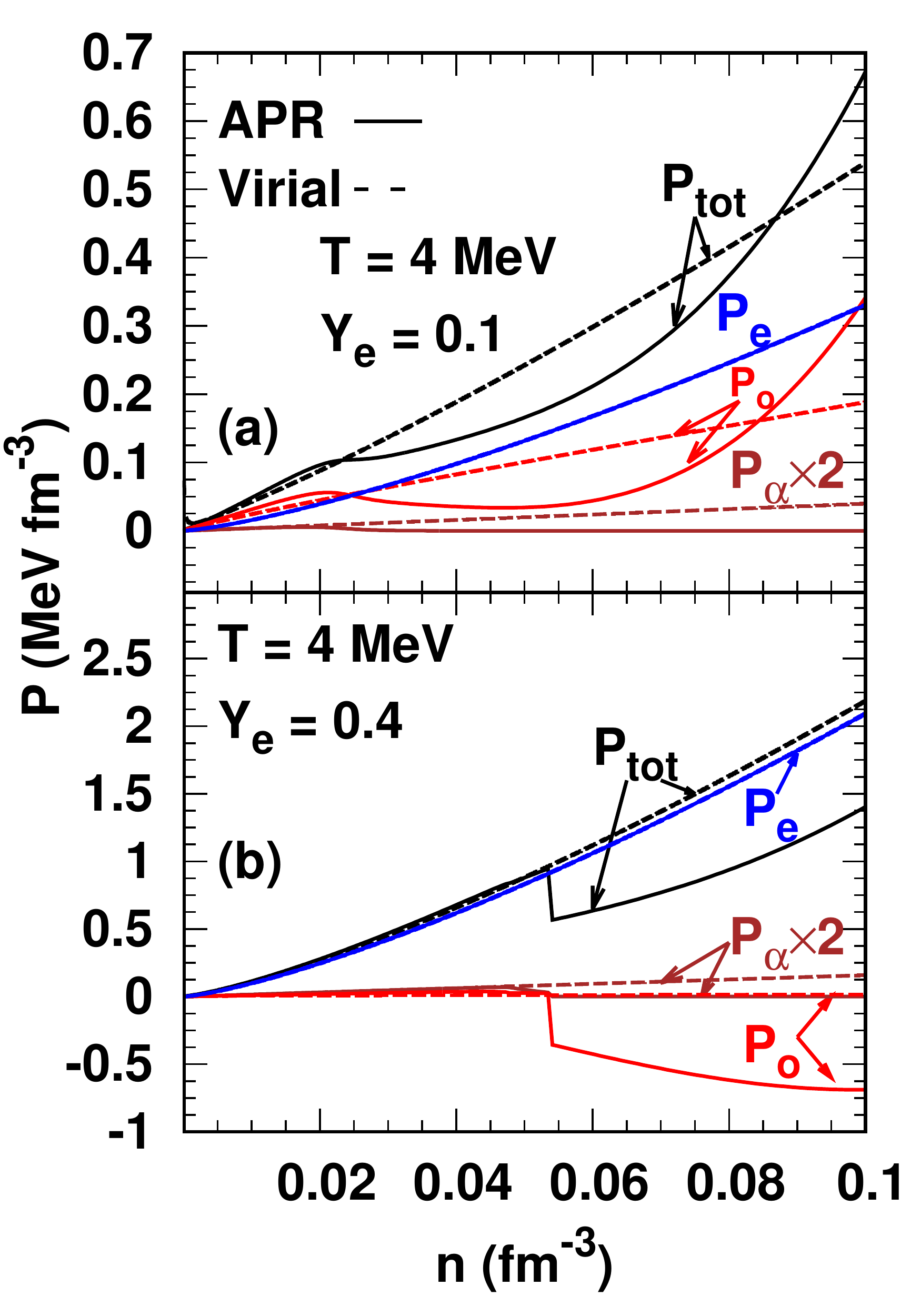}}
\vspace{-0.5cm}
\caption{Pressure vs baryon density in the excluded volume and virial approaches for the indicated temperature and net electron fractions. Low individual contributions have been multiplied by a factor of 2 for clarity.}
\label{fig:Ps}
\end{figure}

\begin{figure}
\centering
\subfloat{\includegraphics[width=0.5\textwidth,keepaspectratio]{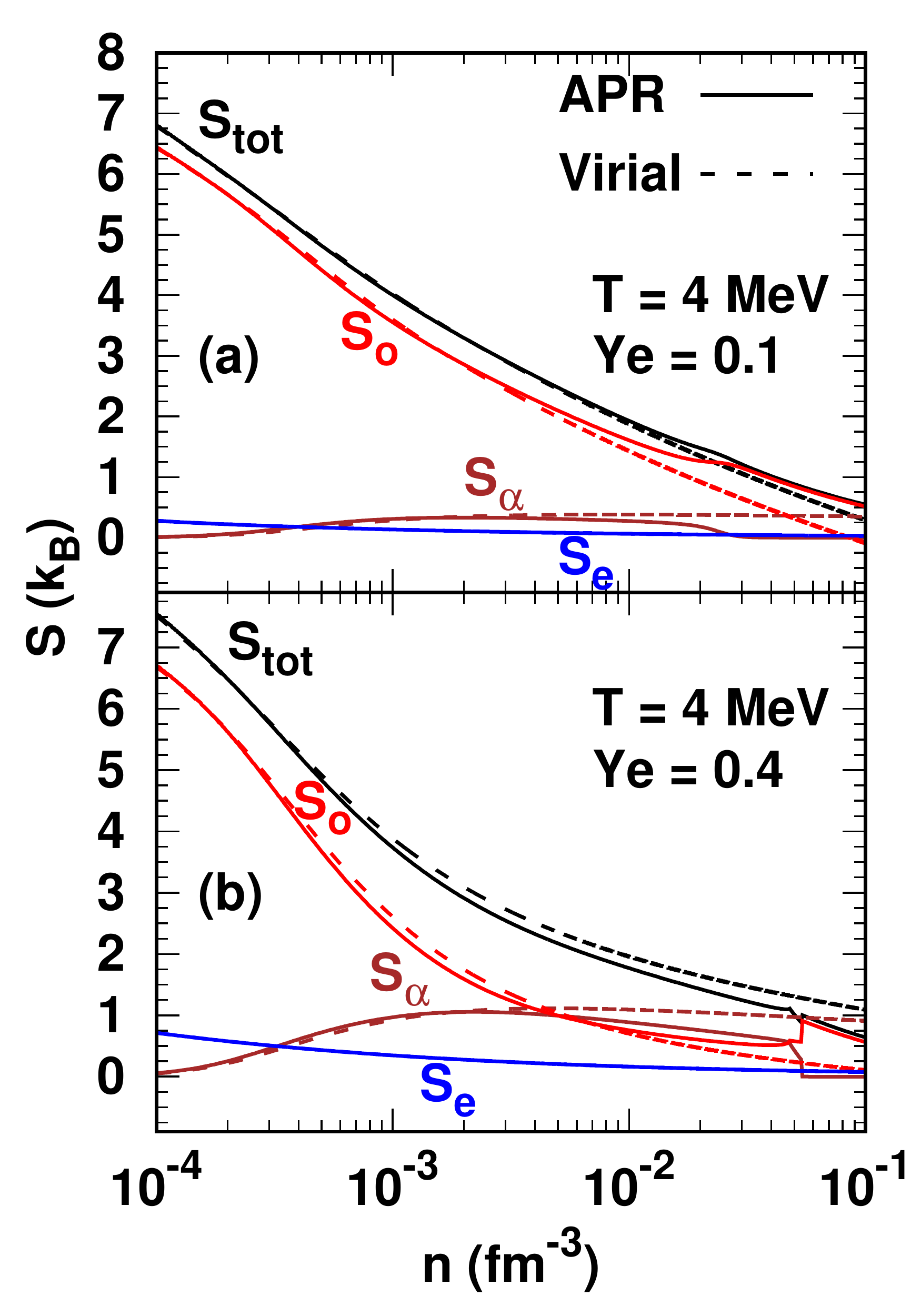}}
\vspace{-0.5cm}
\caption{Same as Fig. \ref{fig:Ps}, but for the entropy per baryon vs baryon number density.}
\label{fig:Ss}
\end{figure}

The entropy per baryon, $S=s/n$, vs $n$ is shown in Figs. \ref{fig:Ss}(a) and (b)  for the same $T$ and $Y_e$'s as in Fig. \ref{fig:Ps}.  The upward trend in the results for $S_{\rm o}$ and $S_e$ is caused by the low values of $n$ in their respective definitions. As with the pressures, the non-monotonic behavior of  $S_{\rm o}$ at the higher end of densities in this figure is caused by the disappearance of $\alpha$-particles in the excluded volume approach.   
As $n_\alpha$ vanishes faster than $n$ for very low $n$, $S_\alpha \rightarrow 0$ as $n \rightarrow 0$.   
In contrast to the pressures, the dominant contribution for $S_{\rm tot}$ arises from the nucleons outside of $\alpha$-particles. 

\begin{figure}
\centering
\subfloat{\includegraphics[width=0.5\textwidth,keepaspectratio]{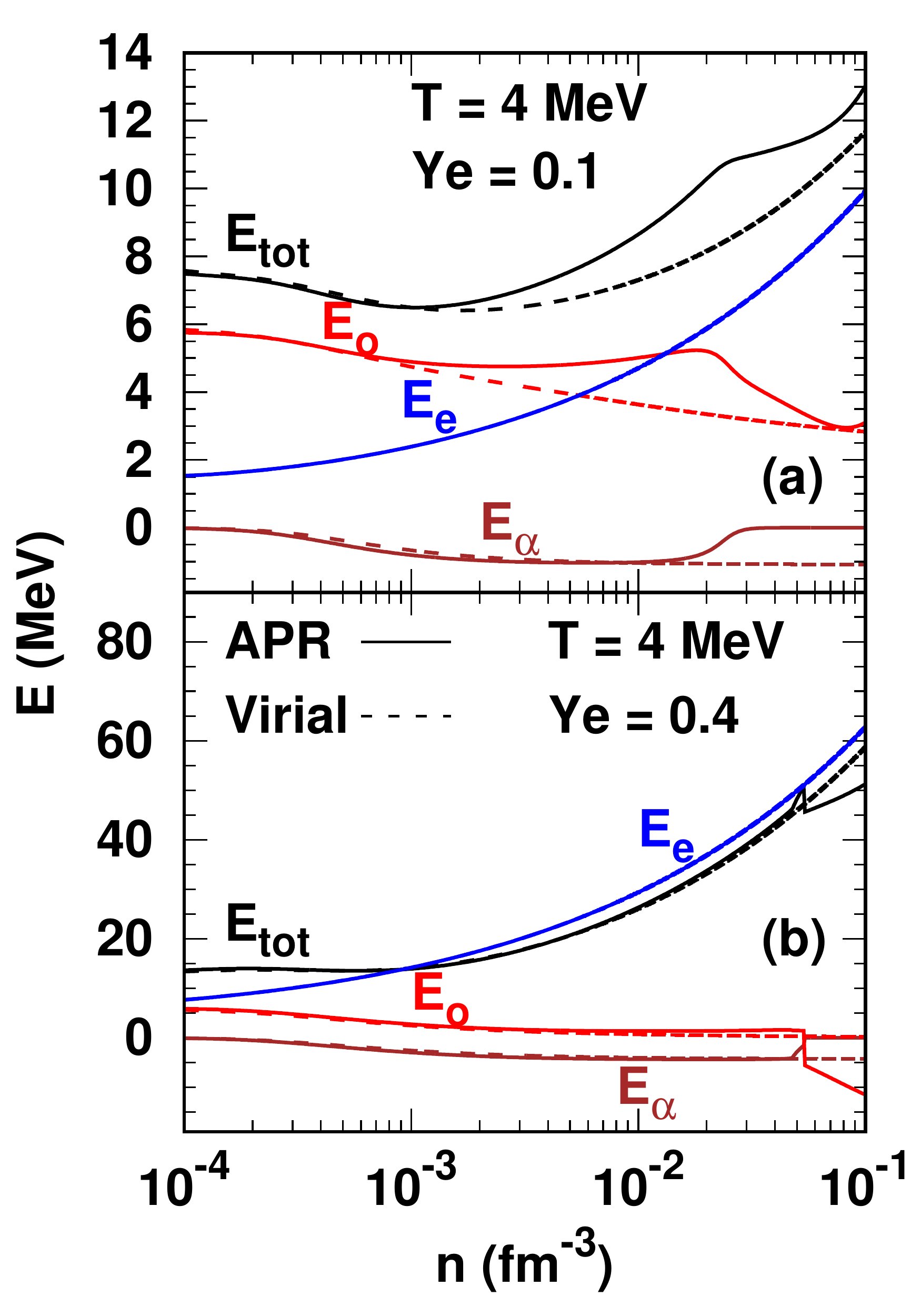}}
\vspace{-0.5cm}
\caption{Same as Fig. \ref{fig:Ps}, but for the energy per baryon vs baryon number density.}
\label{fig:Es}
\end{figure}
For the same values of $T$ and $Y_e$ as in Fig. \ref{fig:Ps}, the energy per baryon, $E=\epsilon/n$, vs $n$ is shown Figs. \ref{fig:Es}(a) and (b).  For both $Y_e$'s, the $\alpha$-particle energies per baryon, $E_\alpha$, in the two approaches are nearly the same at low densities. They differ from each other at the densities for which the excluded volume effects become significant. 
As $n_\alpha$ continues to increase with $n$ in the virial approach, the magnitude of $E_\alpha$ continues to decrease. Note also that $E_\alpha$ remains negative until $T \gtrsim 2B_\alpha/3$. For $Y_e$=0.1, the electron energies supersede those of the nucleons at some density whereas they are the dominant contributions  at $Y_e=0.4$ at all $n$ shown.   In contrast to the virial approach, the non-monotonic behavior of the nucleon energies, $E_{\rm o}$, at $Y_e=0.1$, stemming from the excluded volume approach, is also noteworthy.  \\

\begin{figure}
\centering
\subfloat{\includegraphics[width=0.5\textwidth,keepaspectratio]{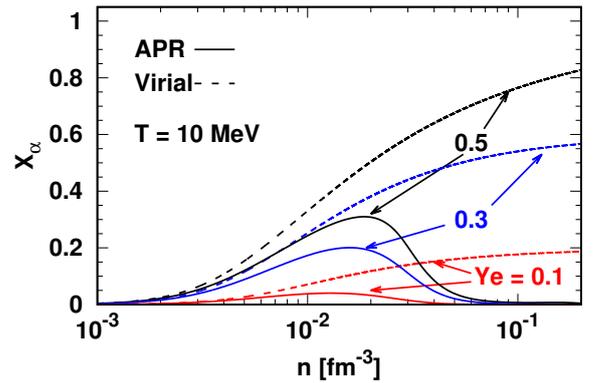}}
\vspace{-0.5cm}
\caption{Same as Fig. \ref{fig:Xas}, but for $T=10$ MeV.}
\label{fig:Xas10}
\end{figure}

In Fig. \ref{fig:Xas10}, the $\alpha$-particle fractions are shown at $T=10$ MeV for representative $Y_e$'s. In contrast to the results at $T=4$ MeV, the dissolution of the $\alpha$-particles is less abrupt in the excluded volume approach. The difference with the results of the virial approach grows as the density increases for all $Y_e$'s.  With more positive charge and baryon number added with the inclusion of additional light nuclear clusters and heavy nuclei, these results will also change correspondingly in both the approaches (see discussion in subsequent sections).

\begin{figure}
\centering
\subfloat{\includegraphics[width=0.5\textwidth,keepaspectratio]{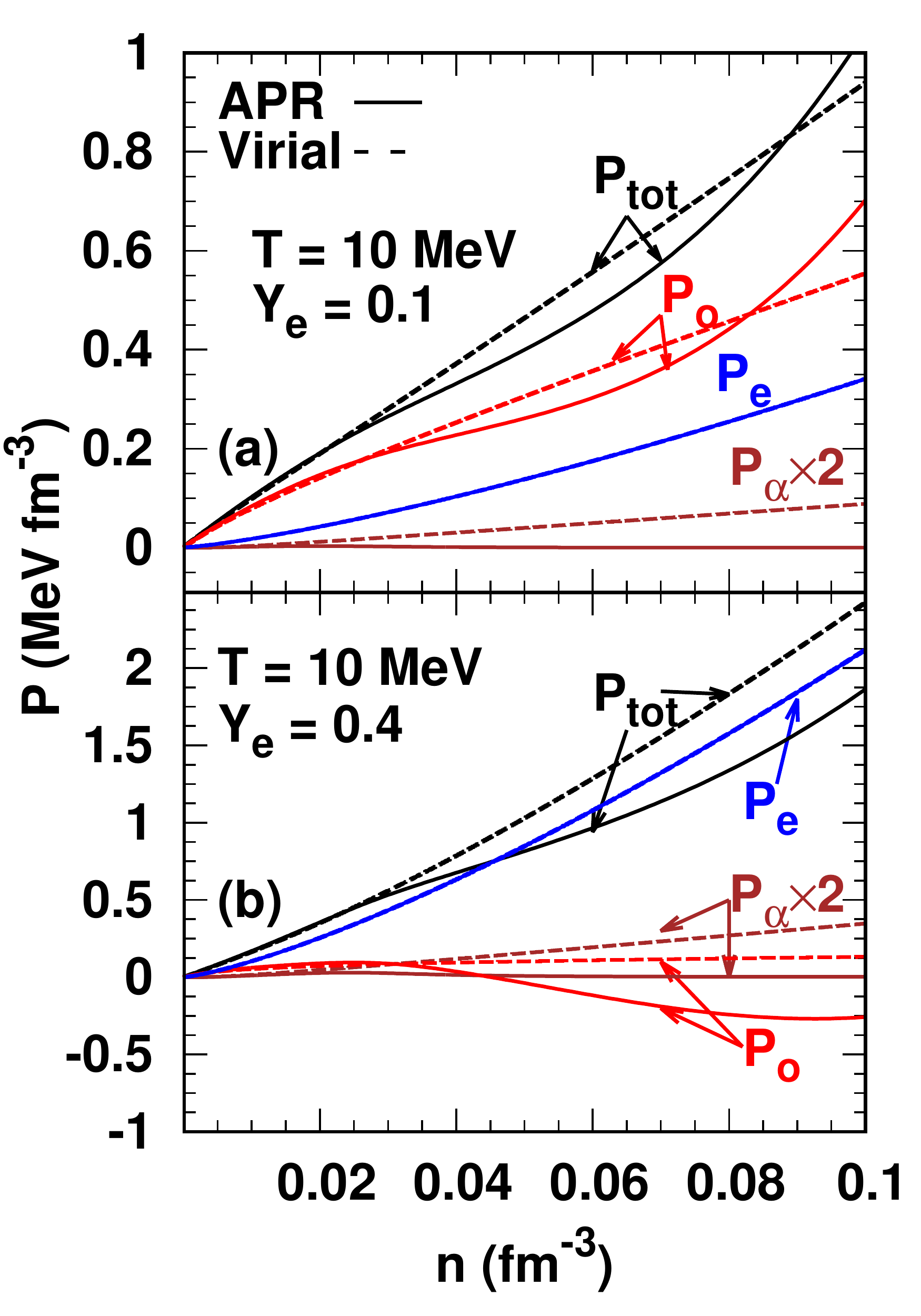}}
\vspace{-0.5cm}
\caption{Same as Fig. \ref{fig:Ps}, but for $T=10$ MeV.}
\label{fig:Ps10}
\end{figure}

Pressures at $T=10$ MeV are shown in Fig. \ref{fig:Ps10}.  Relative to the results in Fig. \ref{fig:Ps} at $T=4$ MeV, the higher thermal content in the pressure of nucleons $P_{\rm o}$ is evident in this figure for both $Y_e$'s. 
Contributions to $P_{\rm o}$ at $Y_e=0.4$ for $n$ approaching $0.1~{\rm fm}^{-3}$ in the excluded volume approach are negative because the temperature is well below the liquid-gas phase transition temperature of $T_c\simeq 17$ MeV for the EOS of APR at this $Y_e$ \cite{CMPL14}.  The  virial   $P_{\rm o}$'s remain positive, albeit very small at $Y_e=0.4$. Contributions from leptons are substantial for both $Y_e$'s.

\subsection*{Results for multiple clusters in the excluded volume  approach}

In this section, we present and discuss results of the excluded volume approach when the low-density phase contains d, $^3$H, $^3$He and $^4$He. For related, but slightly different treatment of the excluded volume approach, {\em cf.} Refs. \cite{Typel16,PT17}. Results in these references are qualitatively similar to those of ours although small quantitative differences exist.  Put together, these results enable  comparisons with results of the virial approach in Refs. \cite{OGHSB07,Arcones08}.

The mass fractions from Eq. (\ref{MMfracs}) are shown in Fig. \ref{fig:MLT4}. In the results shown, the relative fractions of the various species are determined by a combination of the charge and baryon number conservation laws as well as values of the $B_i$'s, $T$, $Y_e$, and $P_{\rm o}$. The density at which an individual species vanishes is primarily controlled by the excluded volume $v_i$ assigned to it. Note that there is some leeway in assigning these values instead of  the geometrical factors adopted here. In principle, one could also use ranges of effective interactions in determining the excluded volume. In addition, the presence of heavy nuclei (to be included later) will also alter the relative concentrations. Our results here are illustrative of the effects of excluded volumes, but the formalism allows for other values  to be used. These remarks apply for results of the state variables shown below as well.  

\begin{figure}
\centering
\subfloat{\includegraphics[width=0.5\textwidth,keepaspectratio]{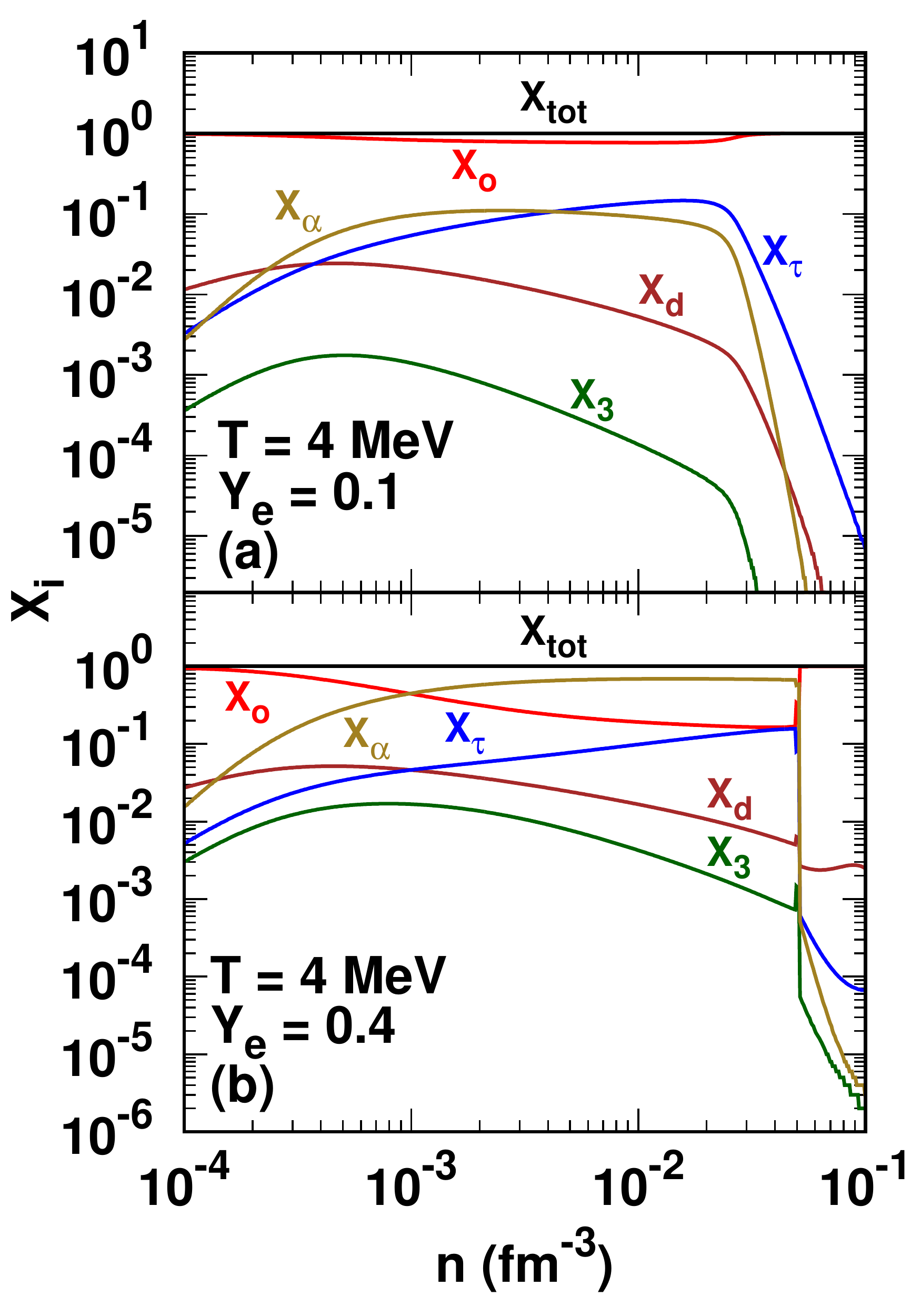}}
\vspace{-0.5cm}
\caption{Mass fractions of light nuclei vs baryon density at the indicated temperature and electron fractions.}
\label{fig:MLT4}       
\end{figure}

Figure \ref{fig:PresT4} shows contributions from light nuclei to the total pressure. As the light nuclei are treated as non-interacting gases, their pressures are given by ideal gas expressions, $P_i=n_i T$, modulo the excluded volume factors in Eq. (\ref{MPres}) which act significantly only when each of the nuclear species is disappearing.   
The temperature being fixed at $T=4$ MeV for this figure, the partial pressures reflect the variation of the individual densities of nuclei with $n$.   Note that contributions from the electrons begin to dominate as $Y_e$ increases. 

\begin{figure}
\centering
\subfloat{\includegraphics[width=0.5\textwidth,keepaspectratio]{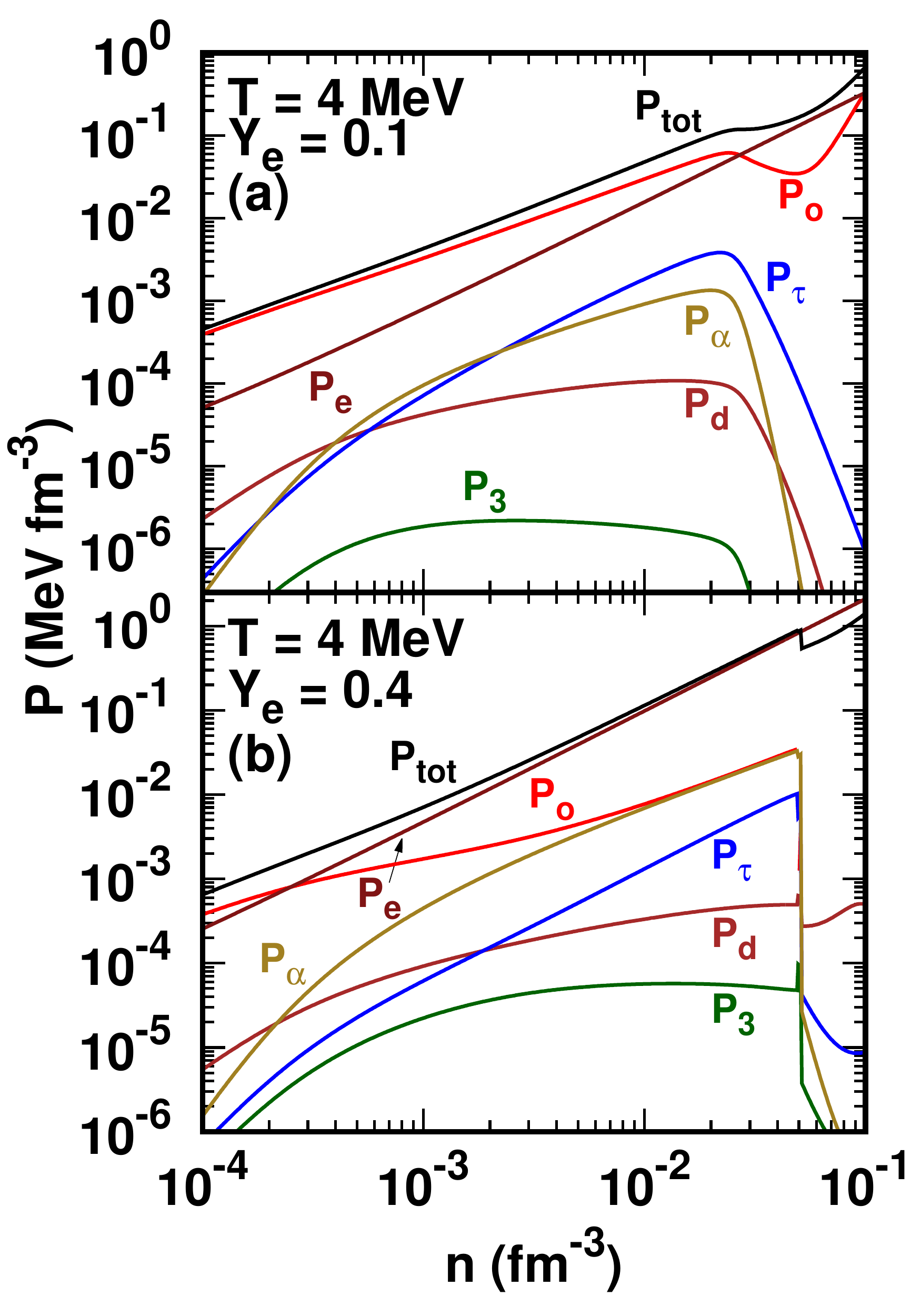}}
\vspace{-0.5cm}
\caption{Contributions from light nuclei to the total pressure vs baryon density at the indicated temperature and electron fractions.}
\label{fig:PresT4}       
\end{figure}

The entropies per baryon of light nuclei and their total are displayed in Fig. \ref{fig:EntT4}.  Note that the predominance of one or the other light nuclear species varies with increasing density with nucleons giving a substantial contribution for both $Y_e$'s shown. 
Contributions from $^3{\rm He}$, d and $e$ are subdominant for both $Y_e$'s. 

\begin{figure}
\centering
\subfloat{\includegraphics[width=0.5\textwidth,keepaspectratio]{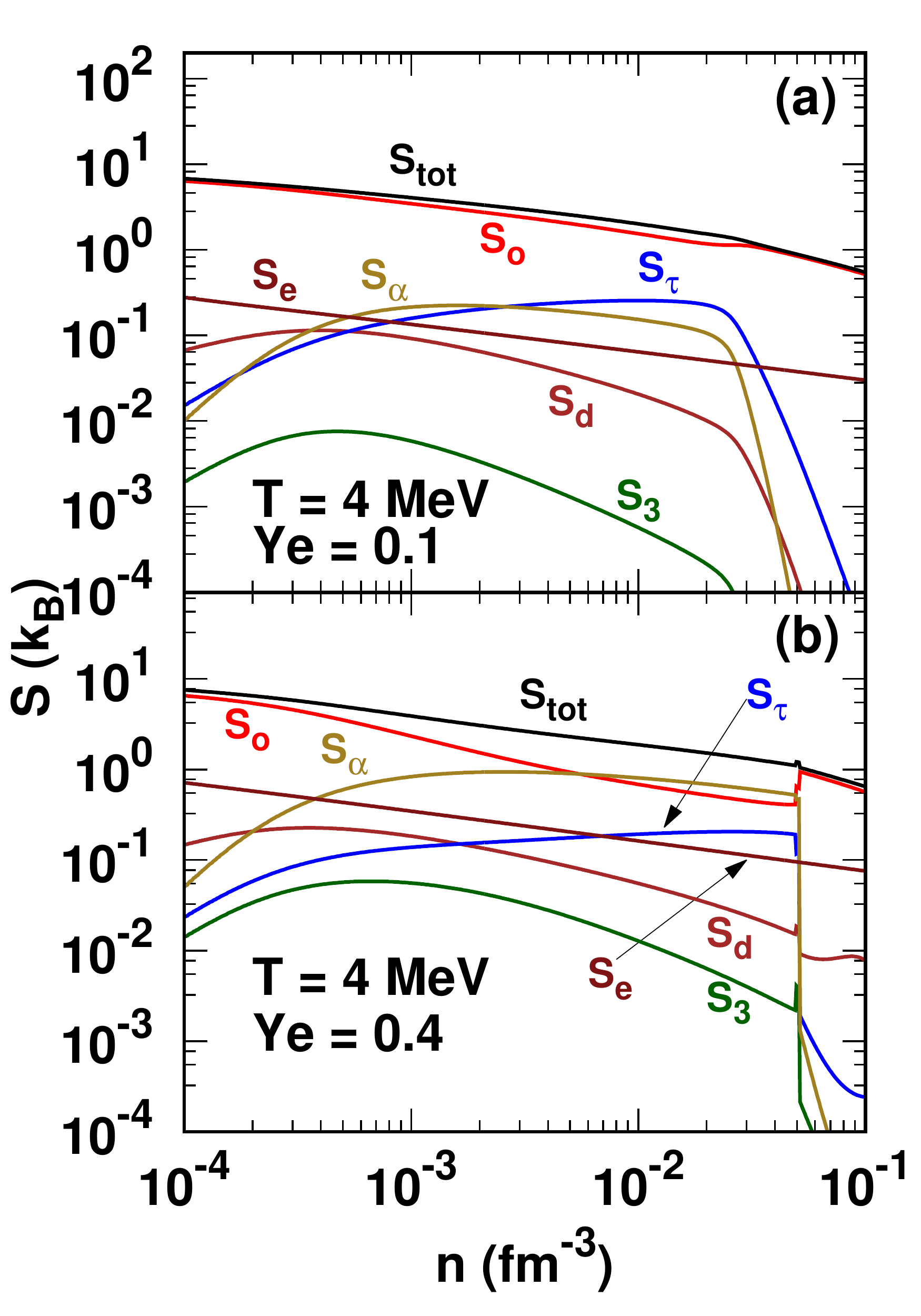}}
\vspace{-0.5cm}
\caption{Contributions from light nuclei to the total entropy per baryon vs baryon density at the indicated temperature and electron fractions. }
\label{fig:EntT4}       
\end{figure}

For the same temperature and $Y_e$'s as in the previous figures, contributions from light nuclei for the total energy per baryon vs density are shown in  Fig. \ref{fig:EnrT4}. Because the B.E. of deuterons is small compared to the temperature, their energies are positive until they disappear. Such is not the case for $^3{\rm H}$, $^3{\rm He}$, and $\alpha$-particles, hence they remain negative until they disappear. Note also that the nucleon energies turn negative as $n$ approaches 0.1 fm$^{-3}$ for $Y_e=0.4$.  
As noted earlier, these results are subject to modifications in the presence of heavy nuclei to be described later. 

\begin{figure}
\centering
\subfloat{\includegraphics[width=0.5\textwidth,keepaspectratio]{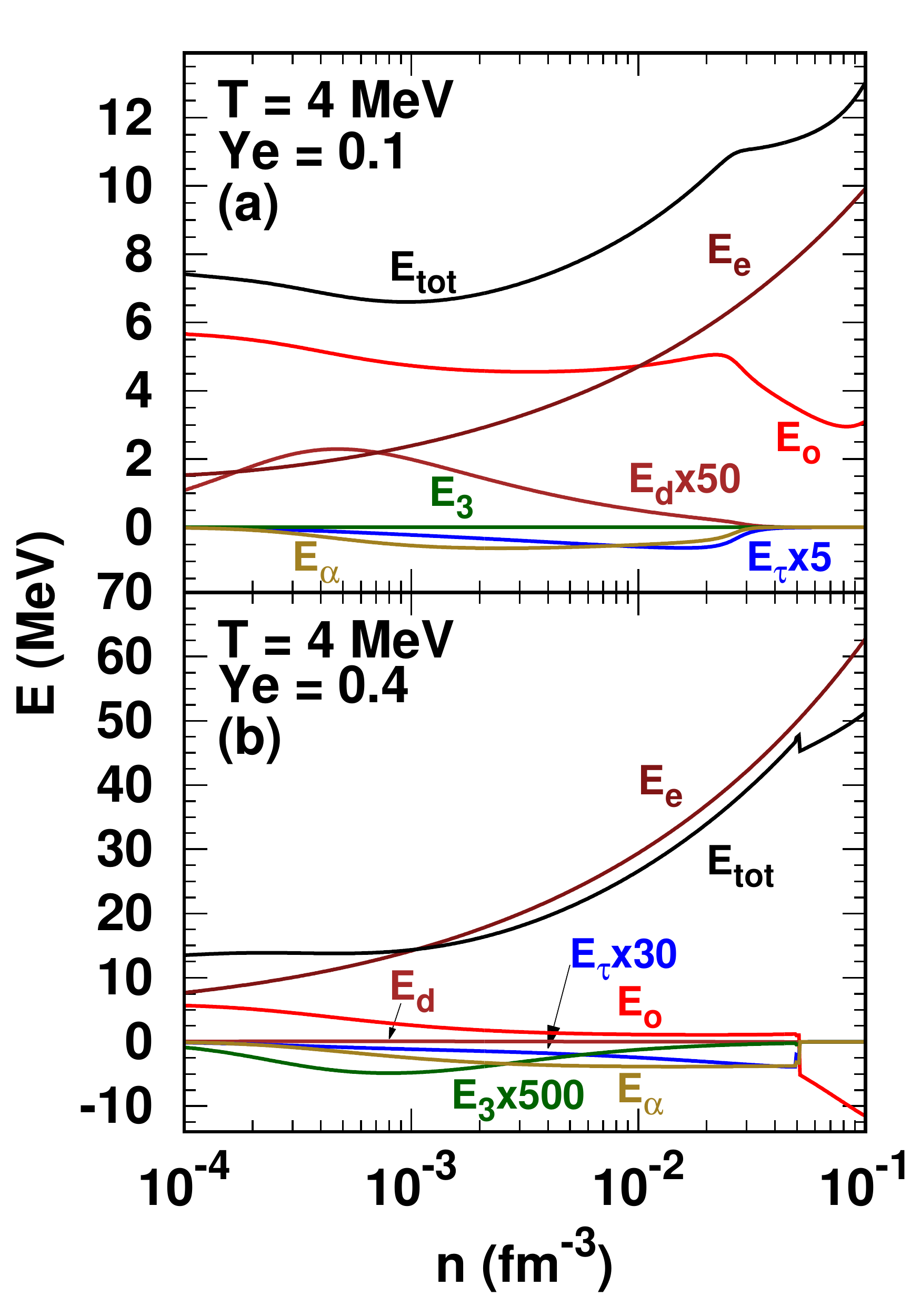}}
\vspace{-0.5cm}
\caption{Contributions from light nuclei to the total energy per baryon vs baryon density.
 Note the large multiplicative factors used in some cases for the sake of clarity.}
\label{fig:EnrT4}       
\end{figure}

\subsection*{Limitations of the excluded volume and virial approaches}

The excluded volume approach accounts only for repulsive interactions which become significant as the density increases. 
As a consequence, light nuclear species disappear at varying densities below  $\sim 0.1~{\rm fm}^{-3}$. In the $(n,Y_e,T)$ region where heavy nuclei are favored, the relative abundances of the light nuclei are also greatly affected \cite{LS91,SHT10}.  The principal drawback of the excluded volume approach is the lack of attractive interactions known to be present from phase shift data, where available.  As inclusion of such effects provides small corrections to the ideal gas state variables,  results of the excluded volume approach will likely not be affected significantly. 

Taking guidance from the available phase shift data, the virial approach, where applicable, includes both attractive and repulsive interactions. In the manner in which interactions are included in this approach, their effects become predominantly attractive. Consequently, the mass fractions of light nuclei continue to increase toward and beyond $n_s$ in the regions of $(n,Y_e,T)$ where heavy nuclei are absent. Although the fugacities of the clusters remain less than unity, such is not the case for nucleons in a wide range of $T$ and $Y_e$.  An illustration of this feature is presented for $npe$ matter in Fig. \ref{fig:zns} where neutron fugacities $z_n$ are shown.  With decreasing proton fraction $x$, the densities at which $z_n$'s exceed unity also decrease. As $z<<1$ is a requirement of the virial approach, caution must be exercised in its use.

From a theoretical standpoint, the region with light nuclear clusters presents the situation of fermion-boson mixtures extensively studied in the context of cold atoms \cite{PS04}.  We have developed an approach based on the mean fields experienced by nucleons and the different light nuclei accounting for  both attractive  and  repulsive hard-core interactions. Results from this approach will be presented in a separate publication.

\begin{figure}
\centering
\subfloat{\includegraphics[width=0.55\textwidth,keepaspectratio]{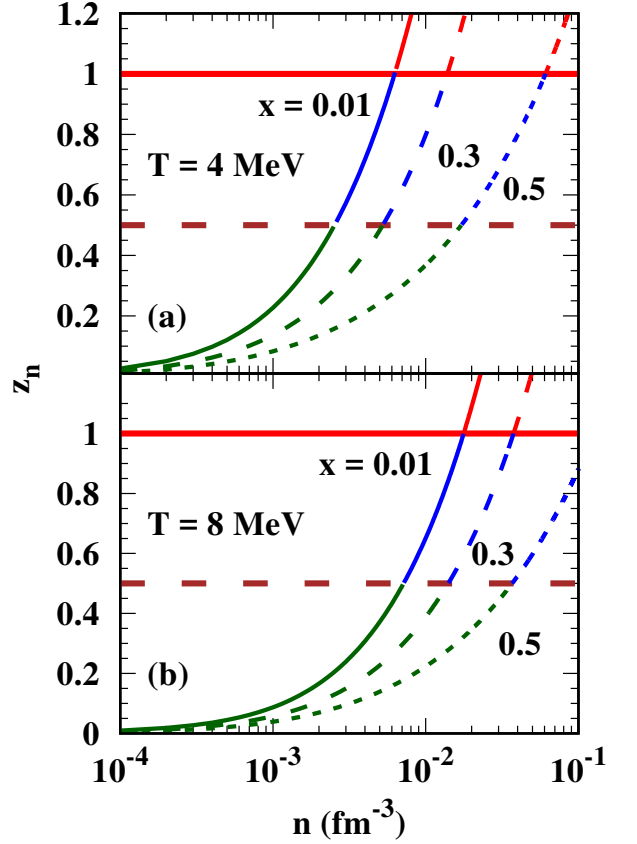}}
\vspace{-0.5cm}
\caption{Neutron fugacity $z_n=\exp(\mu_n/T)$ vs baryon density in $npe$ matter from the virial approach. Values of the temperature $T$ 
and proton fraction $x=n_p/n_b$ are as indicated in the figure. The intersections of the various curves with the horizontal lines at $z_n=1$ indicate the densities beyond which the virial approach loses its validity.}
\label{fig:zns}       
\end{figure}

\subsection*{Inhomogeneous phase with heavy nuclei}

%

%
\begin{figure}
\resizebox{0.5\textwidth}{!}{%
    \includegraphics{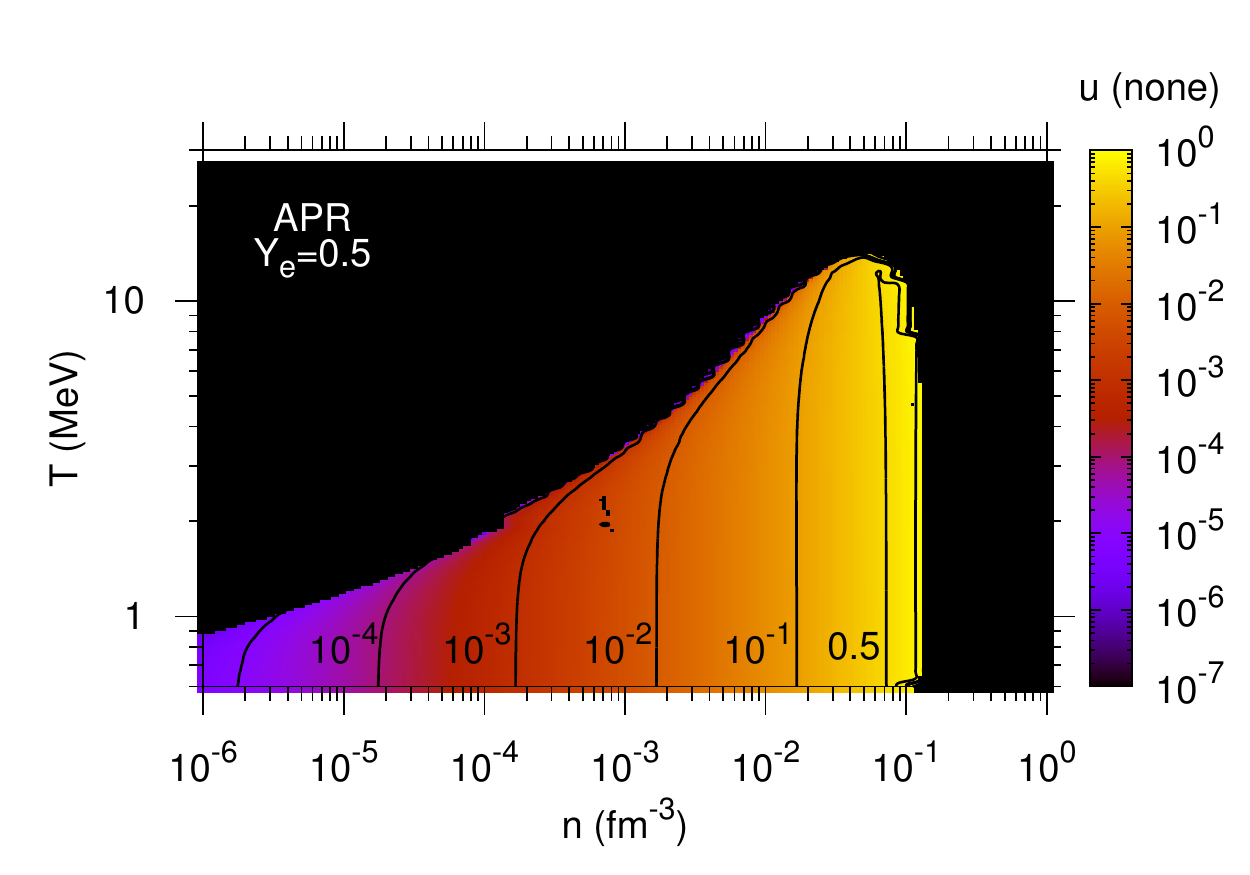}
}
\caption{Contours of the volume fraction $u$ occupied by nuclei at sub-nuclear densities in the $T$-$n$ plane for the EOS of APR. The net electron fraction $Y_e=0.5$. Figure courtesy Brian Muccioli.}

\label{fig:CuAPR}       
\end{figure}

The inhomogeneous phase consisting of heavy nuclei surrounded by a gas of nucleons, light nuclei, leptons and photons occupies a substantial region (phase II in Fig. \ref{fig:Schem}) at sub-nuclear densities $n\lesssim 0.1~{\rm fm}^{-3}$.  
For densities in excess of 
$0.1~{\rm fm}^{-3}$, matter is too dense for nuclei of any type to form and thus consists of uniformly distributed nucleons and leptons (phase III in Fig. \ref{fig:Schem}). For the treatment of heavy nuclei, two main approaches have been adopted in the literature: the single nucleus approximation \cite{LS91,LLPR85} and the full ensemble method \cite{SHT10}. The former approach gives an adequate representation of the thermodynamics of the system \cite{BL84}, while the latter approach is warranted for applications involving neutrino-nucleus and electron-nucleus scattering and absorption processes.

In Fig. \ref{fig:CuAPR}, we show results for the volume fraction $u$ occupied by nuclei for $Y_e=0.5$. The results shown here are for the EOS of APR in the single nucleus approximation of \cite{LS91}. Decreasing the electron fraction from $Y_e=0.5$ reduces the fraction of space occupied by nuclei for a given density $n$ and temperature $T$. This reduction occurs because at low $Y_e$'s, nuclei are unable to maintain a moderate proton fraction ($x_i \simeq 0.3$) and thus fewer nuclei form. 

Observe that while the presence of electrons in the whole system ensures charge neutrality and mechanical stability, electrons are entirely irrelevant in the calculation of $T_c^A$ which is the temperature above which nuclei of mass $A$ dissolve into their constituent nucleons. As such, $T_c$ is a property of matter enclosed in  nuclei and can only depend on the charge fraction $x_i$ inside nuclei.  In turn, $x_i$ is a function of the ambient conditions ($n,Y_e,T$) in which the nuclei are embedded; here $Y_e$ is the net electron fraction of the whole system. For large enough nuclei, matter in their interior can be reasonably well approximated by infinite matter. In this case, $T_c$ is obtained by solving Eq. (9); that is $T_c$ is the liquid-gas phase transition temperature. Note that, in general, $x_i < Y_e$ and therefore $T_c^A < T_c^\infty$. Similar considerations apply to the pasta phase;  the temperature $T_c^p$ above which the pasta phase disappears is much less than $T_c^A$. 
Detailed results for the state variables corresponding to the EOS of APR will be reported separately \cite{SCMP18}.  This work adds to the suite of EOS's based on the Skyrme interaction provided in Ref. \cite{SRO17}.

\subsection*{The supra-nuclear equation of state}

Structural properties, such as the mass, radius, moment of inertia, quadrupolar polarizability (or tidal deformability), etc.,  of a neutron star depend sensitively on the supra-nuclear EOS \cite{LP16}. While the radius  of a normal nucleonic neutron star is primarily determined by the EOS in the region up to $n\sim$ 2$n_s$,  the star's maximum mass depends on the EOS close to  its central density \cite{LP01}.  When additional components such as hyperons, Bose Condensates or quarks are considered  (and present in substantial amounts), the EOS beyond 2$n_s$ can influence both the radius and the maximum mass. The internal composition controls the long-term cooling of neutron stars detected through optical and X-ray thermal emission \cite{PLPS04}.  For the manner in which the supra-nuclear EOS influences other observable properties including gravitational wave emission, neutrino emission from  type-II supernovae, rotation, magnetic properties, etc., of a neutron star, {\it cf.} 
Refs. \cite{LP16,LP07} for an overview. 

There are several {\em sweet and sour points} concerning theoretical attempts to calculate the supra-nuclear EOS. On the sweet 
side, developments in effective field theory have enabled first principle calculations of isospin symmetric and asymmetric matter with systematic corrections to be estimated. On the sour side, continuing beyond 2$n_s$ to encompass the central densities of neutron stars is precluded in these methods because of the perturbative expansion parameter $\Lambda/p$, where $\Lambda$  is a cut-off in momentum $p$, reaching uncomfortable values particularly for PNM for which the Fermi momentum $p_F({\rm PNM}) = 2^{2/3}p_F({\rm SNM})$. For example,  $p_F({\rm PNM})\simeq 336~(533)$ MeV/c for $n_s(2n_s)$, whereas typically $\Lambda \simeq 600$ MeV/c.  Additionally, the error estimates depend on the method employed to impose the cut-off $\Lambda$  which also affects the unitarity of operators when a cut-off is used. To access the EOS beyond 2$n_s$ for inferring structural properties of a neutron star, the approach taken thus far has been to use piece-wise and causal polytropic EOS's beyond 2$n_s$  so that a maximum mass of $\gtrsim 2M_\odot$ can be obtained \cite{HLPS13}. While this approach is adequate and useful for cold neutron stars in a parametric study, the internal composition, finite temperature properties and isospin dependence of the EOS cannot be accessed  with polytropic EOS's.

Phenomenological EOS's based on non-relativistic potential model approaches with contact and finite-range interactions have long been used to explore possible consequences in astrophysical applications by varying the high-density behavior of the EOS.  
The advantage of these models is that calculations are relatively easier than the time-consuming first-principle calculations. However, higher-than-two-body interactions, found necessary to fit  constraints offered by laboratory data on nuclei at near-nuclear densities, render these EOS's acausal at some high density due to the lack of Lorentz invariance in a non-relativistic approach.  Often, the density at which causality is violated lies within the central densities of neutron stars. Although a method to impose causality based on thermodynamical arguments has been known for a while for cold stars \cite{NC73,LPMY90}, it is only recently that a similar method has been devised at finite temperature \cite{CP17}.  A practical way to avoid this problem is to screen repulsive contributions from higher-than-two-body interactions (as they lead to an energy per particle that varies faster than linearly at high density) as in Refs. \cite{BD80,PAL88,Prak97}. 
For any such nonrelativistic potential model, causality is preserved for \textit{all} temperatures/entropies if the inequalities $\frac{c_s(T=0)}{c}\le 1$ and $\frac{4}{9}Q^2+\frac{2n}{3}\frac{dQ}{dn} \le 1$ are \textit{both} satisfied. The quantity $Q$ is related to the nucleon effective mass $m^*$ and its density derivative according to $Q=1-\frac{3m^*}{2n}\frac{dm^*}{dn}$.That is, a full finite-T calculation of the speed of sound is not necessary in order to check whether or not causality is violated.
 These remarks are relevant also to all first-principle dense-matter calculations that use a non-relativistic approach.  \\

Relativistic Dirac-Brueckner-Hartree-Fock \cite{MPA87,Engvik96} and mean field-theoretical \cite{MS96} models and their extensions are inherently Lorentz invariant and thus preserve causality. While the former approach is based on nucleon-nucleon scattering data, in the latter, nucleon-boson coupling \\ strengths  are calibrated in medium at  $n_s$ to reproduce empirical properties of nuclear matter and nuclei. As with their non-relativistic counterparts, several adjustments of many-body forces in medium have been required to obtain 2$M_\odot$ neutron stars with reasonable radii for $\sim 1.4M_\odot$ stars for which observational constraints are beginning to emerge. These approaches, however, do not suffer from the cut-off issues that non-relativistic effective field-theoretical approaches do. \\

Establishing or ruling out the presence of non-nucleonic degrees of freedom in neutron-star matter has proved difficult on both observational and theoretical fronts.  Observationally, a compelling evidence for the presence of exotica in the form of hyperons, Bose condensates or quarks is lacking. On the theoretical side, many studies including the presence of exotica in one form or the other have been conducted. Most  of these studies have been revised with adhoc adjustments concerning strong interactions at high density in view of the discovery of 2$M_\odot$ neutron stars. The consensus since then has been that a large amount of exotica in neutron star interiors is untenable \cite{LP11}. In the case of quarks, the overarching concern is the lack of a non-perturbative treatment of quark matter interactions. This conundrum is likely to remain unless breakthroughs occur on both observational and theoretical fronts.

\section{Thermal effects on the structure of neutron stars}
\label{sec:3}

As Figs. \ref{fig:S} and \ref{fig:T} imply, the entropy and temperature of the post-merger remnant in the merger of binary neutron stars change with the spatial location as well as with time. Consequently, the enclosed mass and the radius of the remnant also change with time.  In reality, such changes are brought about by several physical effects such as thermal effects, neutrino trapping, rotation and magnetic fields etc. all acting at the same time.  To gain a qualitative or semi-quantitative  understanding of how each of these effects affect the masses and radii at a given time, we can study the role of one physical effect at a time while freezing the others. A full dynamical simulation is, however, required for a complete understanding when all of the physical mechanisms act in concert. 

We therefore begin with the role of thermal effects on the structure of a neutron star. As most of the enclosed mass is accumulated from regions above $n_s$ in the star, an analysis based on nearly-degenerate matter offers some insight. Under such conditions, the maximum gravitational mass at finite constant entropy per baryon (throughout the star) can be expressed as \cite{Prak97}
\begin{equation}
M_{\rm max}(S) =  M_{\rm max}(0)~ [ 1 + \lambda S^2 + \cdots ] \,,
\label{MmaxS}
\end{equation}
where the coefficient $\lambda$ is EOS dependent.  Table \ref{tab:MRmaxS} presents physical properties of the maximum mass (gravitational) nucleonic stars for the EOS's chosen in Ref. \cite{Prak97}.   The EOS's labelled BPAL32 and SL32 are non-relativistic potential models, and MRHA and  GM are from mean field theoretical models. The values of $\lambda$ given in Table \ref{tab:MRmaxS} are quite small, $\sim 10^{-2}$. We note that the results in this table correspond to the case when only regions above $n\simeq 0.08~{\rm fm}^{-3}$ contained thermal effects, but not the surface regions below that density.   We have verified that the increasing trend in the maximum mass at finite $S$ is not affected by this omission, but the radius would be  larger when the surface regions are also subject to thermal effects. Results for  EOS's including that  of APR in which the entire star is heated will be reported in a subsequent publication \cite{SCMP18}. 

As in the case of nucleonic stars, thermal effects provide positive pressure at a given baryon density in stars containing hyperons, Bose condensates or quarks as well, and therefore the maximum gravitational mass increases slightly at finite entropy relative to that at zero temperature \cite{Prak97}.  As will be discussed in later sections, other physical effects increase the maximum mass substantially more than thermal effects.



\begin{table}[hbt]
\begin{center}
\caption{Effects of finite entropy on the structure of neutron stars. Results taken from Prakash et al. \cite{Prak97}. 
The various symbols are: $S$, entropy per baryon, $M_{\rm max}$, the maximum mass, $R$, the radius of the maximum mass configuration, $n_c/n_0$, the core density in units of the nuclear equilibrium density, $P_c$, the core pressure, $T_c$, the core temperature, $\lambda$, percentage change in $M_{\rm max}$, and $I$, the moment of inertia corresponding to $M_{\rm max}$. Model designations are as in Ref. \cite{Prak97} (see also text).}
\label{tab:MRmaxS}       
\begin{tabular}{llcclccc}
\hline\noalign{\smallskip}
Model  &  S & $\frac {M_{\rm max}}{M_\odot}$ & $R$ & $\frac {n_c}{n_0}$ & $P_c$ &  $T_c$  & $10^2\lambda$ \\ 
 & & & (km) & & $\left(\frac{ {\rm MeV} } {{\rm fm^{-3}}}\right)$ & (MeV) & \\ \\ \hline 
BPAL32 & 0 & 1.93 & 10.1 & 7.7 & 590.2 & 0 &  \\ 
 & 2 & 1.97 & 10.9 & 6.9 & 482.8 & 71.5 & 0.53 \\ 
SL32 & 0 & 2.1 & 10.6 & 6.8 & 689.9 & 0 & \\ 
& 2 & 2.2 & 11.6 & 5.8 & 532.2 & 103.2 & 1.11 \\ 
MRHA & 0 & 1.86 & 10.6 & 7.3 & 484.9 & 0 & \\ 
& 2 & 1.9 & 11.2 & 6.6 & 419.6 & 58.8 & 0.56 \\ 
GM & 0 & 2.0 & 10.9 & 7.1 & 545.8 & 0 & \\ 
& 2 & 2.04 & 11.6 & 6.4 & 458.2 & 62.6 & 0.47 \\ 
\hline\noalign{\smallskip}
\end{tabular}
\end{center}
\vspace*{2.0cm}
\end{table}

\section{Effects of trapped neutrinos on the structure of neutron stars}
\label{sec:4}

Elusive as they are, the weakly interacting neutrinos can be trapped in matter, albeit transiently, in several astrophysical circumstances. The physical sites of interest include the early universe, core-collapse supernovae, newly born neutron stars, and mergers of binary neutron stars \cite{PLSV01}. For example, the supernova center mean free path for neutrino scattering is $\lambda \approx 2 \times 10^5~(\rm {MeV}/E_\nu)^2$ cm, where $E_\nu$ is the neutron energy in MeV.  Thus, neutrinos with energy 1 MeV or more would be trapped during the evolution of a core-collapse supernova.  Furthermore, at early times in a proto-neutron star's evolution, neutrinos would be  trapped in  matter as well, being unable to propagate on dynamical timescales.   Electron capture reactions, which proceeded due to the increasing density and electron chemical potential, effectively halt as the trapped neutrinos settle into a degenerate Fermi sea and contribute their own  Fermi pressure to the system.  Because neutrinos interact weakly, all of their thermal characteristics are taken as their ideal Fermi gas contributions. 
The chemical potential  of the electron neutrinos, $\mu_{\nu{_e}}$ is related to the net electron-neutrino number per baryon  $Y_{\nu_{e}}$ by $\mu_{\nu_{e}}^3 = 6\pi^2nY_{\nu_{e}}$ (the neutrino mass here being negligible compared to $\mu_{\nu{_e}}$).  
Similar considerations apply to the other flavors of neutrinos as well. 

It is interesting to make some observations regarding the effect of trapped neutrinos on the mass of the star itself to which we turn now.
Under conditions when neutrinos  of lepton  flavor $\ell= e,\mu$ and $\tau$ are trapped in the system, the beta equilibrium condition becomes
\begin{equation}
\mu_ i = b_i\mu_n - q_i(\mu_l - \mu_{\nu_\ell})  \,,
\end{equation}
where $\mu_i$ is the chemical potential of baryon $i$, $b_i$ is its baryon number and $q_i$ is the charge. The chemical potential of the neutron, lepton $\ell$ and neutrino $\nu_\ell$   are denoted by $\mu_n, \mu_\ell$ and $\mu_{\nu_\ell}$, respectively. 
For example, equilibrium under the electron capture reaction $ p + e^- \leftrightarrow  n + \nu_e$ establishes the relation
\begin{equation}
\hat \mu \equiv \mu_n -\mu_p = \mu_e - \mu_{\nu_e} \,,
\end{equation}
allowing the proton chemical potential to be expressed in terms of three independent chemical potentials as 
\begin{equation}
\mu_p = \mu_n - (\mu_e - \mu_{\nu_e}) \,. 
\end{equation}
Analogous relations involving other neutrino flavors, hyperons, kaons, quarks, etc. can be found Ref. \cite{MP96}. 

Because of trapping, the numbers of leptons of each flavor of neutrino
\begin{equation}
Y_{L\ell} = Y_\ell + Y_{\nu_\ell}
\label{YLell}
\end{equation}
are conserved on dynamical time scales.  In the context of core-collapse supernovae, the constraint 
$Y_{L\mu} = Y_\mu + Y_{\nu_\mu} = 0$ can be imposed because no muons are thought to be present when neutrinos become trapped in the gravitational collapse of the white-dwarf core of massive stars. The electron lepton number $Y_{Le} = Y_e + Y_{\nu_e} \simeq 0.4$, the precise value depending on the efficiency of electron capture reactions during the initial collapse stage. Note, however, that in mergers of binary neutron stars, $Y_{L\mu} \neq 0$, as cold catalyzed neutron stars prior to merger would contain some muons above $\sim n_s$.   

Since neutrinos do not carry any charge, the charge neutrality condition remains unaltered from the case in which neutrinos are not trapped.

\begin{table}[hbt]
\caption{Effects of trapped neutrinos on the structure of nucleonic stars. Results taken from Prakash et al. \cite{Prak97}. The various symbols are as in Table \ref{tab:MRmaxS}.}
\label{tab:MRmaxY1}       
\begin{center}
\begin{tabular}{llcclccc}
\hline\noalign{\smallskip}
Model  &  S & $\frac {M_{\rm max}}{M_\odot}$ & $R$ & $\frac {n_c}{n_0}$ & $P_c$ &  $T_c$  & $10^2\lambda$ \\ 
 & & & (km) & & $\left(\frac{ {\rm MeV} } {{\rm fm^{-3}}}\right)$ & (MeV) & \\ \\ 
 \hline
BPAL32 & 0 & 1.86 & 10.1 & 7.6 & 609.6 & 0 &  \\ 
 & 2 & 1.91 & 10.8 & 6.7 & 503.7 & 63.7 & 0.63 \\ 
MRHA & 0 & 1.78 & 10.3 & 7.5 & 514.1 & 0 & \\ 
& 2 & 1.84 & 10.9 & 6.8 & 448.3 & 54.6 & 0.75 \\ 
GM & 0 & 1.94 & 10.5 & 7.4 & 595.8 & 0 & \\ 
& 2 & 1.98 & 11.2 & 6.7 & 496.6 & 59.0 & 0.58 \\ 
\hline\noalign{\smallskip}
\end{tabular}
\end{center}
\vspace*{2.0cm}
\end{table}


The results presented in Table \ref{tab:MRmaxY1} for nucleonic stars correspond to the case when $Y_{Le}=0.4$ and $Y_{L\mu}=0$, i.e., relevant more for core-collapse supernovae and the initial stages of proto-neutron stars than for the post-merger remnants (in which $Y_{L\mu} \neq 0$) of the coalescence of binary neutron stars.  The EOS designations are as in Table \ref{tab:MRmaxS}. Comparing results in Tables \ref{tab:MRmaxY1} and \ref{tab:MRmaxS}, one notices a reduction in the maximum mass of $\sim (0.06-0.08)M_\odot$ depending on the EOS for both $S=0$ and 2 when neutrinos are trapped. This reduction is caused by the electron chemical potential $\mu_e$ being larger in this instance relative to the case of neutrino-free stars. This increase is required by the conservation of $Y_{Le}$ in Eq. (\ref{YLell}) in the presence of neutrinos. Consequently, nucleonic matter becomes more proton rich to maintain charge neutrality, which in turn causes the EOS to become softer relative to the neutrino-free case. 

For the manner in which the maximum mass varies in the case of neutrino-trapped stars containing strangeness-bearing components such as hyperons, kaon condensates or quarks, we refer the reader to Ref. \cite{Prak97} in which pathways of subsidence of such stars to black holes after deleptonization is described.   We note that such pathways would exist only for cases in which a substantial amount of exotica are present (fulfilling,  of course, the current constraint  of $2M_\odot$ cold catalyzed stars).  

As neutrinos diffuse through matter, the lepton numbers will change. The  precise manner in which neutrinos diffuse through the star in time is beyond the scope this work; see, however, Refs. \cite{BL86,KJ95,Pons99,Pons01a,Pons01b,LukeR12} for detailed accounts.

\section{Magnetic effects on neutron star structure}
\label{sec:5}

When magnetic fields have sufficient strength, they influence both the EOS and the structure (through changes in the relevant metric functions) of neutron stars in an intermingled manner. The magnetic field strength needed to dramatically affect neutron structure can be estimated by a dimensional analysis equating the magnetic filed energy $E_B \sim (4\pi R^3/3)(B^2/8\pi)$ with the gravitational binding energy $E_{B.E.} \sim 3GM^2/(5R)$, yielding the so-called virial limit 
\begin{equation}
B \sim 1.4 \times 10^{18} \left(\frac {M}{1.4M_\odot} \right) \left( \frac {R}{10{\rm~ km}} \right)^{-2}~\rm{Gauss} \,,
\end{equation}
where $M$ and $R$ are, respectively, the neutron star mass and radius. 
Magnetic fields quantize the orbital motion (Landau quantization) of charged particles such as electrons, muons and protons in charge neutral and beta-equilibrated matter of neutron stars. The importance of relativistic effects is gauged by the equality of the particle's cyclotron energy $e\hbar B/(mc)$ to its rest mass energy. Although there is nothing critical about this, the magnitudes of the so-called critical fields for the electron, muon and proton are:
\begin{eqnarray}
B_c^e &=& (\hbar c/e)~ \lambdabar_e^{-2} = 4.414 \times 10^{13}~{\rm Gauss} \,, \nonumber \\
B_c^\mu &=& (m_\mu /m_e)^2 B_c^e= 1.755 \times 10^{18}~{\rm Gauss} \,, \nonumber \\
B_c^p &=& (m_p/m_e)^2 B_c^e = 1.487 \times 10^{20}~{\rm Gauss} \,,  
\end{eqnarray}
where $\lambdabar_e = \hbar/(m_e c) \simeq 386$ fm is the reduced Compton wavelength of the electron. When the Fermi energy of the proton becomes significantly affected by $B$, the composition and hence the EOS of matter in beta equilibrium becomes significantly affected, and in general leads to a softening of the EOS for $B^* \equiv B/B_c^e \sim 10^5$ \cite{BPL00}. 

The energy density and pressure from the electromagnetic field are:
\begin{equation}
\epsilon_f = P_f = \frac {B^2}{8\pi} = 4.814 \times 10^{-8}~B^*~\rm{MeV~fm^{-3}} \,.
\end{equation}
Thus, to obtain  a nominal $\epsilon_f=P_f=1~{\rm MeV~fm^{-3}}$, a magnetic field strength of $B^* = 4.56 \times 10^3$ or $B \simeq 2 \times 10^{14}$ Gauss is required.  These values may be contrasted with the pressure of matter in non-magnetic neutron stars that range from 2-4 MeV fm$^{-3}$ at nuclear density to 400-1000 MeV fm$^{-3}$ at the central densities of maximum mass neutron stars  depending on the EOS.  Thus, the field contributions dominate the matter pressure only for $B^*>10^4$ at nuclear densities and for $B^*>10^5$ at the central densities of neutron stars. 

In strong magnetic fields, anomalous magnetic moments of the baryons interact with the magnetic field opposing the softening of the EOS due to Landau quantization \cite{BPL00,BPL02}. For nucleons, 
\begin{eqnarray}
\kappa_p &=& \mu_N \left( \frac {g_p}{2} - 1 \right)~ {\rm for}~ p\,,~{\rm and}\,, \nonumber \\ 
\kappa_n &=& \mu_N \left( \frac {g_n}{2} - 1 \right)  \quad {\rm for}~ n \,,
\end{eqnarray}
where $\mu_N$ is the nuclear magneton, and $g_p=5.58$ and $g_n=-3.82$ are the Lande g-factors for the proton and neutron, respectively. The energy 
\begin{equation}
| \kappa_n + \kappa_p|~ B \simeq 1.67 \times 10^{-5} ~{\rm MeV}
\end{equation}
measures the  changes in the field-free beta equilibrium condition and to the baryon Fermi energies with contributions from the anomalous magnetic moments becoming significant for $B^*>10^5$.  In fact, complete spin polarization of the neutrons occurs when
\begin{eqnarray}
| \kappa_n |~B \sim \frac {(6\pi^2n_n)^{2/3}}{4m_n} \,,
\end{eqnarray}
which at nuclear density leads to $B^* \simeq 1.6 \times 10^5$ or $B \simeq 7.1 \times 10^{18}$ Gauss. 
 Such spin polarization results in an overall stiffening of the EOS (due to the increased degeneracy pressure of neutrons) that counters the softening induced by Landau quantization \cite{BPL00}.  The net effect is to render the effective EOS very close to that of field-free matter. For a summary of additional effects, for fields close to or exceeding $B_c^p$, such as vacuum polarization effects \cite{Sch88} and compositeness of baryons, see Ref. \cite{LP07}.

%
\begin{figure}
\resizebox{0.5\textwidth}{!}{%
    \includegraphics{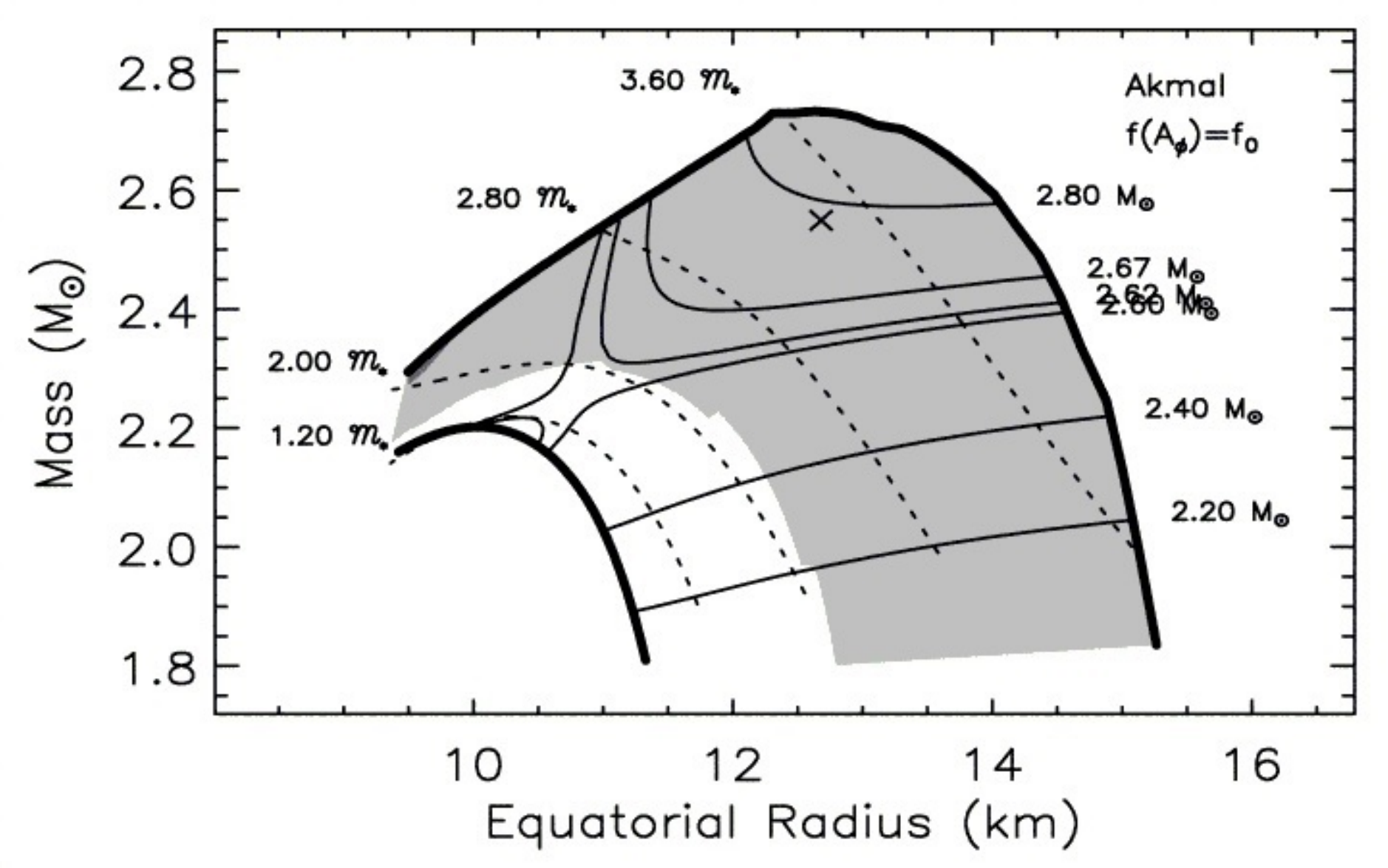}
}
\caption{Mass-equatorial radius plot showing converged solutions attainable with a constant current function for the EOS of APR \cite{APR98}.  The lower heavy curve represents spherical, non magnetized configuration, and the upper heavy curve represents the boundary beyond which solutions appear not to exist.  Lighter solid curves are sequences of constant baryon mass (in $M_\odot$), while dotted curves are sequences of constant magnetic moment ${\cal M}$ (in units of ${\cal M}_*=10^{35}$ Gaussian). 
The cross denotes the maximum mass configuration attainable by uniform rotation.
Figure from Cardall, Prakash and Lattimer \cite{CPL01}.}
\label{fig:MRB}       
\end{figure}

Studies conducted with assumed frozen-in fields, see, e.g., Refs. \cite{BBGN95,BSS00,CPL01} and references therein, offer some insight into the effects of magnetic fields on the structure of neutron stars.  Figure \ref{fig:MRB} shows results from Ref. \cite{CPL01} (for the EOS of APR) in which the limits of hydrostatic equilibrium for axially-symmetric magnetic fields in general relativistic configurations were analyzed assuming a constant current function.  Axially-symmetric magnetic fields provide a centrifugal-like  contribution to the total stress-energy tensor, which flattens an otherwise spherical star. Large enough fields decrease the central (energy) density as the mass is increased eventually compromising the star's stability. As with rotation, magnetic fields allow neutron stars with a particular EOS and baryon number to have larger masses and equatorial radii compared to the field-free case. The maximum mass attainable with a magnetic field governed by a constant current function is noticeably larger than that attained by rotation. 

In Fig. \ref{fig:MRB}, hydrostatically stable configurations (some of which may not be stable to dynamical perturbations) are contained between the heavy solid curves. The lower heavy solid curve is the usual field-free, spherical result for the mass-radius relation. The upper heavy solid curve represents the largest possible stable mass for a given equatorial radius as the internal magnetic field strength is increased. Large axially-symmetric fields tend to yield flattened configurations, and if large enough, shift the maximum densities off-center resulting in toroidal shapes.

The lighter solid curves in Fig.  \ref{fig:MRB} represent constant baryon mass sequences of interest as potential evolutionary paths. 
Note, however, that such paths represent reality only if the current function stays constant over the timescales of magnetic field decay due to Hall drift and ambipolar diffusion. As there is no such guarantee, perhaps the study of several different current functions could shed light on probable evolutionary sequences. The lighter dotted curves in this figure display sequences of constant magnetic moment ${\cal M}$. Unlike for baryon number or mass, there is no principle of ``conservation of magnetic moment'', but over the slow timescales of magnetic field decay, this procedure seems like a plausible opening exploration. 

In the context of simulations of merging neutron stars, several additional caveats apply. The post-merger remnant would be differentially rotating till the time rigid rotation takes over due to effects of poorly understood (artificial) viscosity. The generation and evolution of magnetic fields in the rotating remnant itself is a subject beset with considerable uncertainty and is worthy of further studies.

\section{Rotational effects on neutron star structure}
\label{sec:6}

Rigid rotation increases the maximum mass of a neutron star due to the positive pressure support provided by centrifugal forces. With increasing angular momentum $J$, mass-shedding at the equator occurs, limiting the maximum angular momentum, $J_{\rm max}$, a star can support.   General relativistic instabilities occur slightly before the Keplerian limit, but the latter generally gives a good estimate of rotational instabilities. For uniform rotation, studies  in e.g., Ref. \cite{CST94} have shown that the increase in the maximum gravitational mass can amount to $\sim 20\%$.  
In what follows, we summarize the results of Ref. \cite{BSB18} (see also 
references to earlier works there, and also Ref. \cite{FS13}) in which the behavior of maximum baryon and gravitational masses 
with respect to the Kerr parameter  $a = cJ/(GM^2)$ was studied using a wide class of EOS's and a few select laws for differential rotation. In summary, Ref. \cite{BSB18} finds 
\begin{eqnarray}
\frac {M_B}{M_B^*} &=& 1 + 0.51 \left( \frac {cJ}{GM_B^{*2}} \right)^2 - 0.28  \left( \frac {cJ}{GM_B^{*2}} \right)^4  \\
\frac {M_B}{M_B^*} &=& 0.93~ \frac {M_G}{M_G^*} + 0.07  \\
\frac {M_G}{M_G^*} &=& 1 + 0.29 \left( \frac {cJ}{GM_G^{*2}} \right)^2 - 0.10 \left( \frac {cJ}{M_G^{*2}} \right)^4 \,.
\end{eqnarray}
The notation above is 
$M_G^* := M_{G,max}^{TOV}$,  $M_B^* := M_{B,max}^{TOV}$ are maximum gravitational and baryon masses for non-rotating models with $J=0$, and $M_G$ and $M_B$ are for rotating models with $J\neq 0$. 

Self-bound strange quark stars deviate from above trend yielding  
\begin{eqnarray}
\frac {M_B}{M_B^*} &=& 1 + 0.87 \left( \frac {cJ}{GM_B^{*2}} \right)^2 - 0.60  \left( \frac {cJ}{GM_B^{*2}} \right)^4 \,.
\end{eqnarray}
The conclusion that emerges from these static studies is that uniform (differential) rotation can increase the maximum allowed mass  (before mass shedding) by up to $\sim 20\%~(\geq50\%)$.  
Similar conclusions have been reached in Refs. \cite{BR16,WMR18,RMW18}, albeit with slightly differing numbers in the case of differentially rotating stars with different rotation laws. As noted in Ref. \cite{BSB18}, further analysis by extracting realistic rotation laws from dynamical simulations (including magnetic effects) is warranted. 

\section{Conclusions}
\label{sec:7}

Simulations of the merger of binary neutron stars require the EOS of dense matter over wider ranges of density and temperature than do those of core-collapse supernovae and protoneutron stars. This requirement stems mainly from  the  compression and mass of the post-merger remnant achieved in a merger event which are larger than those in the latter cases. Although advances have been made in dense matter theory, many sore points remain some of which have been pointed out in this work.  The gravitational and baryon masses of the post-merger remnants are also influenced by effects of composition, temperature, neutrino trapping, magnetic fields and rotation, the latter differential for short times and rigid thereafter. To gain physical insights, we have provided brief reviews of earlier works studying these effects considering each of them to act separately.  In dynamical simulations of mergers, however, all of these effects would be acting simultaneously  and evolving with time.

New elements of our work here are (1) a comparison of excluded volume and virial approaches for the np$\alpha$ system using the EOS of APR for interacting nucleons, and, (2) extension of the excluded volume approach to include additional light nuclei such as d, $^3$H, and $^3$He at sub-nuclear densities along the lines of Refs. \cite{LS91,LLPR85}.  

The principal difference between the excluded volume and virial approaches for the np$\alpha$ system is that the mass fraction of the $\alpha$-particle vanishes for $n\lesssim 0.1~{\rm fm}^{-3}$ in the former case (due to excluded volume effects) whereas it continues to rise for the latter up to and beyond nuclear densities. As a result, the excluded volume total pressure exhibits a non-monotonic behavior with density for all electron fractions unlike in the virial approach.  For the same reason, similar features are also seen in the total energy per baryon. In both cases, the dominant contribution to the entropy per baryon comes from nucleons outside of $\alpha$-particles for all electron fractions. The origin of the differences between the two approaches is that the excluded volume approach accounts only for repulsive interactions whereas interactions in the virial approach are predominantly attractive.  In addition, the requirement that fugacities be less than unity is not met for nucleonic matter; the density at which the violation occurs decreasing with proton fraction.  For densities, temperatures and electron fractions for which heavy nuclei would be present, results  from both approaches are similar although quantitative differences exist. 

Results from our extension of the excluded volume approach to include light nuclei in addition to the $\alpha$-particle enable comparisons to be made with related, but slightly different approach of Refs. \cite{Typel16,PT17}.  We defer such a comparison to a future work.  We find that  variation of an order in magnitude in the excluded volumes does not result in a big variation of results when multiple clusters are present.   What does, in fact, determine the relative mass fractions of light nuclei are the respective binding energies and, to a lesser extent, the charge fraction at which the calculation is performed; for example, in neutron rich matter, the concentrations of $^3$H and $^3$He will be somewhat enhanced in comparison to the symmetric nuclear matter case.
The EOS in this density region would also be relevant to intermediate energy heavy-ion collisions in which abundances of these nuclei are measured.  One must note, however, the contributions from electrons, present in astrophysical situations, would be absent in this case.  A worthwhile future task would be analyses of fermion-boson mixtures at sub-nuclear densities  using effective field theoretical (EFT) techniques.

EFT approaches have enabled first-principle calculations of isospin symmetric and asymmetric nucleonic matter put to $\sim 2n_s$ with systematic error estimates associated with perturbation theory, the treatment of three-body interactions and dependencies on cut-off procedures.  For densities beyond $\sim 2n_s$, extrapolations using causal polytropic EOS's have been used to examine the ranges of masses and radii consistent with the requirement of obtaining $2M_\odot$ neutron stars. This latter  procedure precludes extension of the EOS to finite temperatures; furthermore, knowledge of the compositional dependence of the EOS required in dynamical simulations is lost.  Other approaches, including phenomenological EOS's based on non-relativistic potential models, relativistic Dirac-Brueckner-Hartree-Fock and mean field theoretical models (and its extensions) do not suffer from cut-off issues. However, several adjustments to many-body forces in such treatments have been required since the discovery of  $2M_\odot$ neutron stars.
Concerning the possible presence of exotica (hyperons, Bose condensates and quarks) in neutron star interiors, the $2M_\odot$ constraint places restrictions in that a significant amount of such matter is disfavored.

Studies of static configurations in which thermal effects, neutrino trapping,  assumed magnetic fields and rotation are present on an individual basis have revealed that the maximum gravitational and baryon masses are affected to varying degrees. The maximum increase in these masses occurs for rotation nearing the mass-shedding (Keplerian) limit. 
For rigid (differential) rotation, the increase in the maximum can amount to $20\%(\gtrsim 50\%)$.  Magnetic fields have a comparable effect only for fields close to or in excess of $10^{18}$ Gauss. In both of these cases, the star would be deformed. Changes in maximum masses due to thermal effects are comparatively smaller, of order a per cent, which is also the case when neutrinos are trapped. To better understand results of dynamical simulations in which all of these effects would be acting concurrently, study of static configurations in which these effects are combined one after the other with a wider choice of magnetic fields and differential rotation laws than employed so far would be greatly helpful.    

\section{Acknowledgements}
\label{ack}
Research support from the U.S. DOE grant. No. DE-FG02-93ER-40756 is gratefully acknowledged. We thank David Radice, Jim Lattimer, Andreas Bauswein and Luciano Rezzola for helpful communications. This work benefited from discussions at the 2018 INT-JINA Symposium on ``First multi-messenger observation of a neutron star merger and its implications for nuclear physics'' supported by the National Science Foundation under Grant No. PHY-1430152 (JINA Center for the Evolution of the Elements) as also from discussions at the 2018 N3AS collaboration meeting of the
``Research Hub for Fundamental Symmetries, Neutrinos, and Applications to Nuclear Astrophysics'' supported by the
National Science Foundation, Grant PHY-1630782, and the Heising-Simons Foundation, Grant 2017-228.

\bibliographystyle{h-physrev3}
\bibliography{epjaref_prakash}

\end{document}